\documentclass[journal]{IEEEtran}
\usepackage[sort,compress]{cite}
\usepackage{graphicx}
\usepackage{hyperref}
\usepackage{amsmath}
\usepackage{amssymb}
\usepackage{xcolor}
\usepackage{enumerate}
\usepackage[all]{xy}

\def \R{{\mathbb R}}
\def \N{{\mathbb N}}

\def \C{\mathcal{C}}

\def \H{\mathcal{H}}
\def \KL{\mathcal{KL}}

\def \ki{\mathcal{K}_{\infty}}
\def\KLL{\mathcal{KLL}}
\def \U{\mathcal{U}}
\def \S{\mathcal{S}}
\def \SW{\mathcal{SW}}
\def \T{\mathcal{T}}
\def \comp{\,{\scriptstyle\circ}\,}

\def\K{\mathcal{K}}
\def\Ki{\K_{\infty}}
 \def\mer{\hfill $\circ$}
 
 \def\sign{\mathrm{sign}}

 \def\I{\mathcal{I}}
\def\vphi{\boldsymbol{\varphi}}
\def\valpha{\boldsymbol{\alpha}}
\def\id{\mathrm{id}}
\def\blambda{\boldsymbol{\lambda}}

\DeclareMathOperator*{\esssup}{ess.sup}

\newtheorem{teo}{Theorem}[section]
\newtheorem{lema}[teo]{Lemma}
\newtheorem{cor}[teo]{Corollary}
\newtheorem{prop}[teo]{Proposition}

\newtheorem{defin}[teo]{Definition}
\newtheorem{as}{Assumption}
\newtheorem{rem}{Remark}
\newtheorem{example}{Example}

\newcommand{\qed}{$\hfill\blacksquare$}

\begin{document}

\title{Uniform Input-to-State Stability for\\ Switched and Time-Varying Impulsive Systems}

%

\author{Jos\'e L. Mancilla-Aguilar and Hernan Haimovich%
  \thanks{J.L. Mancilla-Aguilar is with Departamento de Matem\'atica, Instituto Tecnol\'ogico de Buenos Aires, Avda. Madero 399, Buenos Aires, Argentina. (e-mail: \texttt{jmancill@itba.edu.ar).}}%
  \thanks{H. Haimovich is with the International Center for Information and Systems Science (CIFASIS), CONICET-UNR, Ocampo y Esmeralda, 2000 Rosario,
    Argentina. (e-mail: \texttt{haimovich@cifasis-conicet.gov.ar})}
     \thanks{Work partially supported by ANPCyT grant PICT 2018-1385, Argentina.}}%

\maketitle

\begin{abstract}
  We provide a Lyapunov-function-based method for establishing different types of uniform input-to-state stability (ISS) for time-varying impulsive systems. The method generalizes to impulsive systems with inputs the well-established philosophy of assessing the stability of a system by reducing the problem to that of the stability of a scalar system given by the evolution of the Lyapunov function on the system trajectories. This reduction is performed in such a way so that the resulting scalar system has no inputs. Novel sufficient conditions for ISS are provided, which generalize existing results for time-invariant and time-varying, switched and nonswitched, impulsive and nonimpulsive systems in several directions.
\end{abstract}


\section{Introduction}
\label{sec:introduction}

Impulsive systems are dynamical systems whose state evolves continuously most of the time but may exhibit jumps (discontinuities) at isolated time instants \cite{lakbai_book89}. The continuous evolution of the state (i.e. between jumps) is governed by ordinary differential equations. The time instants when jumps occur are part of the impulsive system definition and the after-jump value of the state vector is governed by a static (i.e. not differential) equation. If the impulsive system has inputs, these may affect both the flow (i.e. the continuous evolution) and the jump equations \cite{yanpen_mcs18}. 

When investigating stability of a system, it is important to characterize the effects of external inputs. The concept of input-to-state stability (ISS), originally introduced for continuous-time systems in \cite{sontag_tac89}, has proved useful in this regard. ISS were subsequently extended and studied for other classes of systems: time-varying systems \cite{edwlin_cdc00}, discrete-time systems \cite{jiawan_auto01}, switched systems \cite{mangar_scl01}, hybrid systems \cite{caitee_scl09} and infinite dimensional systems \cite{dasmir_mcss13,mirwir_tac17}.

The stability of an impulsive system is related to the interplay between the flow and jump equations. For example, in one impulsive system the flow equation may be destabilizing and the jump equation stabilizing, and in another the converse situation may hold. Hence, in either situation the frequency of occurrence of jumps may be decisive as far as stability is concerned. Sufficient conditions for ISS based on Lyapunov-type functions and on the frequency of jump occurrence have been derived in \cite{heslib_auto08}. The results of \cite{heslib_auto08} apply to impulsive systems where both the flow and jump equations are time-invariant. Note, however, that even in the case where neither equation depends explicitly on time, the impulsive system is not time-invariant due to the fact that the impulse times are fixed and part of the system definition.

Since the appearance of \cite{heslib_auto08}, many works have addressed the stability of impulsive systems with inputs from ISS-related standpoints \cite{chezhe_auto09,liuliu_auto11,liuliu_scl12,daskos_nahs12,dasmir_siamjco13,liudou_scl14,lizha_auto17,dasfek_nahs17,lili_scl18,penden_scl18,peng_ietcta18}. Among these works, we can find results for linear time-varying flow and jump equations \cite{liudou_scl14}, for nonlinear time-invariant flow equation \cite{dasmir_siamjco13,dasfek_nahs17,  fekbaj_ecc18, fekbaj_auto19, fekbaj_ecc19}, for impulsive systems with time delays \cite{chezhe_auto09,daskos_nahs12,lizha_auto17,penden_scl18,peng_ietcta18}, and for impulsive systems involving switching \cite{liuliu_auto11,liuliu_scl12,lili_scl18}. In addition, some results for hybrid systems may also be applicable to impulsive systems \cite{libnes_tac14,miryan_auto18,liuhil_amc18}.

The Lyapunov-function conditions in a majority of these results (except for {\cite{liuliu_scl12, dasmir_siamjco13,fekbaj_ecc18, fekbaj_auto19, fekbaj_ecc19}}), however, resemble exponential-type conditions: during flows, the time-derivative along any trajectory is not greater than some coefficient times the value of the Lyapunov function; after jumps, the value of the Lyapunov function is not greater than another coefficient times the value immediately before the jump. Some generalization of the form of these conditions appears in \cite{penden_scl18,peng_ietcta18,ninhe_is18}, where some of the coefficients recently mentioned can be time-varying. To the best of our knowledge, \cite{liuliu_scl12, dasmir_siamjco13,fekbaj_ecc18, fekbaj_auto19, fekbaj_ecc19} are the only works among those previously mentioned that provide results employing a qualitatively more general form for the conditions on the Lyapunov function in the sense that the conditions are genuinely nonlinear on the Lyapunov function. 

As is well-known, the ISS property imposes a bound on the state trajectory comprising a decaying-to-zero term whose amplitude depends on the initial state value, and an input-magnitude-dependent term. As already explained in \cite{heslib_auto08}, the decaying term in the ISS definition employed for impulsive systems decays as elapsed time progresses but is insensitive to the occurrence of jumps. This is in contrast with hybrid systems \cite{caitee_cdc05,caitee_scl09} where the decaying term also decreases when a jump occurs.

In this paper, we consider two ISS notions, namely weak and strong ISS. The decaying term in the former is insensitive to jumps whereas that of the latter causes additional decay with each jump. The weak ISS property is the one considered in most of the literature of impulsive systems, while strong ISS is a standard stability concept for hybrid systems (see \cite{caitee_scl09} and \cite{liuhil_amc18}). The concept of strong ISS gives a more accurate description of the behaviour of the trajectories of the system, especially when the impulse-time sequence has no dwell or average dwell time. In addition, for such a stronger stability property it is possible to show that ISS implies integral ISS \cite{haiman_auto19b_arxiv}. Another reason for considering this strong stability property is that it is robust \cite{goesan_book12}, whereas the weak counterpart is not \cite{haiman_auto19c_arxiv}.

We provide a method for establishing the uniform ISS of families of impulsive systems, based on Lyapunov-type functions (Theorem~\ref{thm:main}). 
In essence, the method mimics the already standard philosophy for nonimpulsive systems of reducing the problem to the assessment of the stability of systems given by the evolution of the Lyapunov functions on the system trajectories. Our construction is such that these comparison-type systems have two salient advantages: they are scalar and have no inputs. Moreover, the conditions imposed on the Lyapunov-type functions, and hence on the resulting comparison-type systems, are sufficiently general so as to allow assessing uniform ISS of impulsive systems over classes of impulse-time sequences as well as over families of systems whose system functions belong to some prescribed sets, as in the case of switched impulsive systems. The ISS results we give in this paper contain several existing ones for both impulsive and nonimpulsive, switched and nonswitched systems as particular cases. To allow even greater generality, our results are given in the (time-varying) two-measure framework introduced by Movchan in \cite{mov_JAMM60} for nonimpulsive systems without inputs (see the book \cite{lakliu_book93} for a general treatment of the stability in two measures and the papers \cite{kelwir_nolcos13} and \cite{trakel_cdc15} for extensions to system with inputs).

Based on the given method, we derive several sufficient conditions for ISS. Specifically, Theorem~\ref{teo:arbitrary} applies to impulsive systems where the continuous part is stabilizing. This theorem generalizes and strengthens existing results for nonimpulsive time-invariant and time-varying systems in \cite{sonwan_scl95,ninghe_scl12,cheyan_tac17}. Theorem~\ref{thm:avdt-radt} and its Corollary~\ref{cor:adt-radt} provide results when the impulse-time sequences satisfy average or reverse-average dwell-time conditions and generalize/strengthen results in \cite{heslib_auto08,ninhe_is18,lili_mcs19}. Theorem~\ref{thm:sciss1} applies when the impulse-time sequences have minimum or maximum dwell times and generalize and strengthen results in \cite{dasmir_siamjco13}. We also give two sets of sufficient conditions especially suited to switched impulsive systems in Theorems~\ref{thm:avdt-radtsw} and~\ref{thm:sw}, which generalize and strengthen results in \cite{vucha_auto07,liuliu_scl12}.

This section ends with a brief explanation of the notation employed. In Section~\ref{sec:stab-impuls-syst}, we describe the type of systems considered as well as the stability concepts employed and their interrelationships. Our main result, namely the Lyapunov function-based method, is given in Section~\ref{sec:main-result}. Sections~\ref{sec:suff-cond-iss} and~\ref{sec:switched} derive sufficient conditions for ISS, all based on our main result. The results in Section~\ref{sec:switched} are especially suited to switching impulsive systems. Section~\ref{sec:technical-proofs} contains some technical or lengthy proofs. Conclusions are given in Section~\ref{sec:conclusions}.

\textbf{Notation.} $\N$, $\R$, $\R_{>0}$ and $\R_{\ge 0}$ denote the natural numbers, reals, positive reals and nonnegative reals, respectively. $|x|$ denotes the Euclidean norm of $x \in \R^p$. We write $\alpha\in\K$ if $\alpha:\R_{\ge 0} \to \R_{\ge 0}$ is continuous, strictly increasing and $\alpha(0)=0$, and $\alpha\in\Ki$ if, in addition, $\alpha$ is unbounded. We write $\beta\in\KL$ if $\beta:\R_{\ge 0}\times \R_{\ge 0}\to \R_{\ge 0}$, $\beta(\cdot,t)\in\Ki$ for any $t\ge 0$ and, for any fixed $r\ge 0$, $\beta(r,t)$ monotonically decreases to zero as $t\to \infty$. For any function $h:I\subset \R\to \R^p$, $h(t^-)$ and $h(t^+)$ denote, respectively, the left and right limits of $h$ at $t\in \R$, when they exist and are finite. For every $n\in\N$ and $r\ge 0$, we define the closed ball $B_r^n := \{x\in\R^n : |x| \le r\}$. Without risk of confusion, if $\gamma = \{\tau_k\}_{k=1}^N$, then $\gamma$ can be interpreted as both the sequence $\{\tau_k\}_{k=1}^N$ and the set $\{\tau_k: k\in\N, 1 \le k < N+1\}$. For a set $S\subset \R$, $|S|$ denotes the Lebesgue measure of $S$. A function $h : D  \to \R^n$, with $D\subset \R \times \R^n$ an open or closed set, is said to be a Carath\'eodory function if $h(t,\xi)$ is measurable in $t$ for fixed $\xi$, continuous in $\xi$ for fixed $t$, and for every compact set $K \subset D$, there exists an integrable function $m_K(t)$ such that $|h(t,\xi)| \le m_K(t)$ for all $(t,\xi)\in K$ (see \cite[Sec.~I.5]{hale_book80}).

\section{Stability of Impulsive Systems with Inputs}
\label{sec:stab-impuls-syst}

\subsection{Impulsive systems with inputs}
\label{sec:prel-defs}

Consider the time-varying impulsive system with inputs $\Sigma$ defined by the equations
 \begin{subequations}
   \label{eq:is}
   \begin{align}
     \label{eq:is-ct}
     \dot{x}(t) &=f(t,x(t),u(t)),\phantom{x(t^-)+g^-}\quad\text{for } t\notin \gamma,    \displaybreak[0] \\
     \label{eq:is-st}
     x(t) &=x(t^-)+g(t,x(t^-),u(t)),\phantom{f} \quad\text{for } t\in \gamma,
   \end{align}
 \end{subequations}
where $t\ge 0$, the state variable $x(t)\in \R^n$, the input variable $u(t)\in \R^m$, $f$ and $g$ are functions from $\R_{\ge 0}\times \R^n\times \R^m$ to $\R^n$, and $\gamma=\{\tau_k\}_{k=1}^{N} \subset (0,\infty)$, with $N$ finite or $N=\infty$ is the impulse-time sequence. We shall refer to $f$ and to (\ref{eq:is-ct}) as, respectively, the flow map and the flow equation and to $g$ and to (\ref{eq:is-st}) as, respectively, the jump map and the jump equation. By ``input'', we mean a Lebesgue measurable and locally essentially bounded function $u:[0,\infty)\to \R^m$; we denote by $\U$ the set of all the inputs. As is usual for impulsive systems, we only consider impulse-time sequences $\gamma=\{\tau_k\}_{k=1}^N$ that are strictly increasing and have no finite limit points, i.e. $\lim_{k\to \infty}\tau_k=\infty$ when the sequence is infinite; we employ $\Gamma$ to denote the set of all such impulse-time sequences. For any sequence $\gamma = \{ \tau_{k} \}^N_{k=1} \in \Gamma$ we define for convenience $\tau_0=0$ and $\tau_{N+1}=\infty$ when $N$ is finite; nevertheless, $\tau_0$ is never an impulse time, because $\gamma \subset (0,\infty)$ by definition.

We assume that 
for each input $u\in \U$ the map $f_u(t,\xi):=f(t,\xi,u(t))$ is a Carath\'eodory function and hence the (local) existence of solutions of the differential equation $\dot x(t)=f(t,x(t),u(t))$ is ensured (see \cite[Thm.~I.5.1]{hale_book80}).

The impulsive system $\Sigma$ is completely determined by the sequence of impulse times $\gamma$ and the flow and jump maps $f$ and $g$. Hence, we write $\Sigma=(\gamma,f,g)$.  Given $\gamma \in \Gamma$ and an interval $I\subset [0,\infty)$, we define $n^\gamma_I$ as the number of elements of $\gamma$ that lie in the interval $I$: 
\begin{align}
  n^\gamma_I &:= \# \big[ \gamma\cap I \big].
\end{align}

A solution of $\Sigma=(\gamma,f,g)$ corresponding to an initial time $t_0\ge 0$, an initial state $x_0\in \R^n$ and an input $u\in \U$ is a 
function $x:[t_0,T_x)\to \R^n$ such that: 
\begin{enumerate}[i)]
\item $x(t_0)=x_0$; 
\item $x$ is locally absolutely continuous on each interval $J = [t_1,t_2) \subset [t_0,T_x)$ without points of $\gamma$ in its interior, and $\dot{x}(t)=f(t,x(t),u(t))$ for almost all $t\in J$; and \label{item:solflow}
\item for all $t\in \gamma \cap (t_0,T_x)$, the left limit $x(t^-)$ exists and is finite, and it happens that $x(t) = x(t^-)+g(t,x(t^-),u(t))$.\label{item:soljump}
\end{enumerate}
Note that \ref{item:solflow}) implies that for all $t\in [t_0,T_x)$, $x(t)=x(t^+)$, i.e. $x$ is right-continuous at $t$.

The solution $x$ is said to be maximally defined if no other solution $y:[t_0,T_y)$ satisfies $y(t) = x(t)$ for all $t\in [t_0,T_x)$ and has $T_y > T_x$. 
We will use $\T_{\Sigma}(t_0,x_0,u)$ to denote the set of maximally defined solutions of $\Sigma$ corresponding to initial time $t_0$, initial state $x_0$ and input $u$. 
%
\begin{rem}
  Note that even if $t_0 \in \gamma$, any solution $x\in\T_{\Sigma}(t_0,x_0,u)$ begins its evolution by ``flowing'' and not by ``jumping''. This is because in item~\ref{item:soljump}) above, the time instants where jumps occur are those in $\gamma \cap (t_0,T_x)$.\mer
\end{rem}

\subsection{Families of impulsive systems}
\label{sec:families}

Often one is interested in determining whether some stability property holds not just for a single impulse-time sequence $\gamma\in\Gamma$ but also uniformly for some family $\S \subset \Gamma$. For example, the family $\S$ could contain all those impulse-time sequences having some minimum, maximum or average dwell time. Another situation of interest is to determine if some stability property holds not just for a single pair of functions $(f,g)$ but also for all pairs $(f,g)$ 
belonging to some given set $\mathcal{F}$.  
To take into account these and other situations, we consider a parametrized family $\Sigma_\Lambda := \{\Sigma_{\lambda}=(\gamma_{\lambda},f_{\lambda},g_{\lambda})\}_{\lambda \in \Lambda}$ of impulsive systems with inputs, where $\Lambda$ is an index set (i.e. an arbitrary nonempty set). For example, if we are interested in studying stability properties of systems modelled by (\ref{eq:is}) which hold uniformly over a class $\S\subset \Gamma$, then we set $\S$ as the index set, and consider the parametrized family of systems $\{\Sigma_{\gamma}=(\gamma,f,g)\}_{\gamma \in \S}$. By taking as index set $\Lambda = \mathcal{F}$ and considering the family $\{\Sigma_{(f,g)}=(\gamma,f,g)\}_{(f,g) \in \Lambda}$ we can handle the other mentioned situation. Another interesting situation we can handle in our framework is that of switched impulsive  systems (see for instance \cite{liuliu_scl12}). This will be explained in Section~\ref{sec:switched}.

\subsection{Stability definitions}
\label{sec:stab-defs}

Stability notions for systems with inputs that are uniform with respect to initial time, such as uniform ISS, 
bound the state trajectory in relation to initial state, elapsed time and input. In the context of impulsive systems, the input can be interpreted as having both a continuous-time and an impulsive component. From (\ref{eq:is-st}) one observes that the values of $u$ at the instants $t\in\gamma$ may instantaneously affect the state trajectory. For this reason, input bounds suitable for the required stability properties have to account for the instantaneous values $u(t)$ at $t\in\gamma$. Given an input $u \in \U$, an impulse-time sequence $\gamma \in \Gamma$ and an interval $I\subset \R_{\ge 0}$, we thus define
\begin{align}
  \| u_I \|_{\gamma} &:= \max\left\{ \esssup_{t\in I} |u(t)| , \sup_{t\in \gamma\cap I} |u(t)| \right\}, \label{eq:iss-norm}
\end{align}
When $I=[0,\infty)$ we simply write $\|u\|_{\gamma}$. This definition is in agreement with that employed in \cite{caitee_cdc05,caitee_scl09} in the context of hybrid systems.

To perform stability analysis in terms of two measures (see \cite{liuliu_scl12,chalib_jco06}), we consider the set $\mathcal{H}$ of functions $h:\R_{\ge 0}\times \R^n\to \R_{\ge 0}$ and define the following stability notions. The reader not familiar with stability in terms of two measures can still get a fair idea of our results by considering only the standard ISS property. However, the use of the two-measure framework allows far greater generality.
\begin{defin}
   \label{def:stab}
Let $h^o, h\in \mathcal{H}$. We say that the parametrized family $\Sigma_\Lambda = \{\Sigma_{\lambda} = (\gamma_{\lambda},f_{\lambda},g_{\lambda})\}_{\lambda \in \Lambda}$ of impulsive systems with inputs is  
\begin{enumerate}[a)]
\item weakly $(h^o,h)$-ISS if there exist $\beta \in \KL$ and $\rho \in \ki$ such that for all $\lambda\in \Lambda$, $t_0\ge 0$, $x_0\in \R^n$, $u\in \U$ and $x\in \T_{\Sigma_{\lambda}}(t_0,x_0,u)$, it happens that for all $t\in [t_0,T_x)$
  \begin{align}
    \label{eq:cwiss}
    h(t,x(t)) &\le \beta \left (h^o(t_0,x_0),t-t_0\right )
    +\rho(\|u_{(t_0,t]}\|_{\gamma_{\lambda}});
  \end{align}
\item strongly $(h^o,h)$-ISS if there exist $\beta \in \KL$ and $\rho \in \ki$ such that for all $\lambda\in \Lambda$, $t_0\ge 0$, $x_0\in \R^n$, $u\in \U$ and $x\in \T_{\Sigma_{\lambda}}(t_0,x_0,u)$, it happens that for all $t\in [t_0,T_x)$
  \begin{multline}
    \label{eq:ciss}
     h(t,x(t)) \le \beta \left (h^o(t_0,x_0),t-t_0+n^{\gamma_{\lambda}}_{(t_0,t]}\right )
    \\+\rho(\|u_{(t_0,t]}\|_{\gamma_{\lambda}}).
  \end{multline}
\end{enumerate}
\end{defin}
By suitable selection of $h^0$ and $h$, one can recover the definitions of different stability properties usually considered in the analysis of both impulsive and nonimpulsive systems. For example, with $h^0(t,x)=h(t,x)=|x|$, the weak $(h^o,h)$-ISS property becomes the standard ISS property considered in the literature of systems with inputs. So, we say that $\Sigma_{\Lambda}$ is weakly or strongly ISS when it is, respectively, weakly or strongly $(h^o,h)$-ISS with $h^0(t,x)=h(t,x)=|x|$. By considering, in addition, that the set where the inputs take values is $\R^0:=\{0\}$, then the standard definition of global uniform asymptotic stability (GUAS) for systems without inputs is recovered. By taking $h^0(t,x)=|x|$ we obtain an extension of the input-to-output stability property (IOS) studied in \cite{sonwan_siamjco00}; see \cite{liuliu_scl12} for more examples. 
\begin{rem}
\label{rem:caus-ISS}
Due to causality and the Markov property, equivalent definitions of $(h^o,h)$-ISS are obtained if $\|u_{(t_0,t]}\|_{\gamma_{\lambda}}$ is replaced by $\|u\|_{\gamma_{\lambda}}$ in (\ref{eq:cwiss}) or (\ref{eq:ciss}). Note that we do not require the solutions of (\ref{eq:is}) to be defined for all $t\ge t_0$ in the definitions of the different stability properties. In general, the $(h^o,h)$-ISS property does not by itself imply the existence of the solution $x(t)$ for all times $t\ge t_0$, 
since $x(t)$ may be unbounded on the finite interval $[t_0,T_x)$ while $h(t,x(t))$ may remain bounded on that interval. An additional condition that ensures existence of the solutions for all times $t\ge t_0$ is the following: for every $M\ge 0$ and every finite interval $J\subset \R_{\ge 0}$, there exists $M'\ge 0$ such that $h(t,x) \le M$ and $t\in J$ imply $|x|\le M'$. \mer
\end{rem}

All the properties in Definition~\ref{def:stab} are uniform with respect to both initial time $t_0$ and the different systems within the family $\Sigma_\Lambda$. The decaying term in a weak property is insensitive to jumps, whereas that of a strong property forces an additional decay whenever a jump occurs. The weak ISS property is the one considered in most of the literature on impulsive systems with inputs, 
 whereas strong ISS is in agreement with the ISS property for hybrid systems as in \cite{libnes_tac14}. Strong 
$(h^0,h)$-ISS from Definition~\ref{def:stab} could be defined equivalently replacing $\beta \left (h^o(t_0,x_0),t-t_0+n^{\gamma_{\lambda}}_{(t_0,t]}\right )$ by $\tilde\beta\left( h^o(t_0,x(t_0)),t-t_0,n^\gamma_{(t_0,t]} \right)$ with $\tilde\beta \in \KLL$\footnote{We write $\tilde\beta\in\KLL$ if $\beta:\R_{\ge 0}\times \R_{\ge 0} \times \R_{\ge 0}\to \R_{\ge 0}$, $\tilde\beta(\cdot,s,\cdot)\in\KL$ and $\tilde\beta(\cdot,\cdot,s)\in\KL$ for every $s\ge 0$.}. The latter form, with $h(t,x) = h^o(t,x) = |x|$, is the one employed in \cite{caitee_cdc05,caitee_scl09}. The equivalence between these is explained in footnote~2 on p.~1397 of \cite{libnes_tac14} and based on Lemma~6.1 of \cite{caitee_tac07}.

\subsection{Relationship between stability properties}
\label{sec:relat-betw-stab}

Since $\beta(r,s+t) \le \beta(r,s)$ for every $(r,s,t)\in\R_{\ge 0}^3$ and every $\beta\in\KL$, it is clear that 
strong $(h^o,h)$-ISS implies weak $(h^o,h)$-ISS. The following property will be useful in establishing the converse implication.
\begin{defin}
  \label{def:UIB}
  Consider a set $\S\subset\Gamma$ of impulse-time sequences. We say that $\S$ is uniformly incrementally bounded (UIB) if there exists a continuous and nondecreasing function $\phi : \R_{\ge  0} \to \R_{\ge 0}$ so that $n^{\gamma}_{(t_0,t]} \le \phi(t-t_0)$ for every $\gamma\in \S$ and all $t>t_0\ge 0$. 
\end{defin}
Note that sequences corresponding to minimum or average dwell time are UIB.
The following example shows that, however, not every $\gamma\in\Gamma$ is UIB.
\begin{example}
  \label{ex:infns}
  Consider the sequence $\gamma = \{\tau_k\}_{k=1}^\infty$ with $\tau_1 = 1$ and $\tau_{k+1} = \tau_k + 1/(k+1)$. Note that $\gamma$ is a strictly increasing sequence and $\lim_{k\to\infty} \tau_k = \sum_{k=1}^\infty (1/k) = \infty$. Then $\gamma$ has no finite limit points and hence $\gamma\in\Gamma$. However, $\lim_{k\to\infty} \tau_{k+1}-\tau_k = \lim_{k\to\infty} 1/(k+1) = 0$, and hence consecutive elements of $\gamma$ occur closer together as time increases. Then, if we consider the interval $(t_0,t_0+1]$ and the number of elements of $\gamma$ that fall within the latter interval, namely $n^\gamma_{(t_0,t_0+1]}$, it follows that $\lim_{t_0\to\infty} n^\sigma_{(t_0,t_0+1]} = \infty$. This shows that $\gamma$ is not UIB.\mer 
\end{example}

Under UIB, weak and strong $(h^0,h)$-ISS become equivalent.
\begin{prop} 
  \label{prop:equivalence}
  Let $\Sigma_\Lambda = \{\Sigma_{\lambda}=(\gamma_{\lambda},f_{\lambda},g_{\lambda})\}_{\lambda \in \Lambda}$ be a parametrized family of impulsive systems with inputs and let $h,h^o \in \mathcal{H}$. If $\Gamma_\Lambda = \{\gamma_{\lambda} : \lambda\in \Lambda \}$ is UIB, then $\Sigma_{\Lambda}$ is strongly $(h^0,h)$-ISS if and only if it is weakly $(h^0,h)$-ISS.
\end{prop}
The proof of Proposition \ref{prop:equivalence} is provided in Section~\ref{sec:proof-prop-equiv}. The relationships between the stability properties considered can be summarized in the following chart.
\begin{displaymath}
  \xymatrix{
  \text{strong ($h^0,h$)-ISS} \ar@/^/@2{->}[r] & \ar@/^/@2{->}[l]^{\text{UIB}} \text{weak ($h^0,h$)-ISS}
}
\end{displaymath}

\section{Main Result}
\label{sec:main-result}

In this section we give 
a result for establishing the weak or strong $(h^0,h)$-ISS of a parametrized family of impulsive systems with inputs. This result involves the existence of a family of Lyapunov-like functions. In order to relax regularity conditions on the latter, especially in the case of impulsive switched systems, we introduce the following classes of functions. Given an impulse-time sequence $\gamma =\{\tau_k\}_{k=1}^N \in \Gamma$, we say that $V:\R_{\ge 0}\times \R^n\to \R$ belongs to class $\mathcal{V}(\gamma)$, and write $V\in\mathcal{V}(\gamma)$, if 
\begin{enumerate}[i)]
 \item $V$ is locally Lipschitz on\footnote{Recall that $\tau_0=0$ and that $\tau_{N+1}=\infty$ when $N$ is finite.} $[\tau_k,\tau_{k+1})\times \R^n$ for $0 \le k< N+1$;
 \item for each $\tau \in \gamma$ and $\xi\in \R^n$, $\lim_{(t,\zeta)\to (\tau^-,\xi)}V(t,\zeta)=\lim_{t\to \tau^-}V(t,\xi)=:V(\tau^-,\xi)$.
\end{enumerate}
Note that if $V\in\mathcal{V}(\gamma)$, then $V(\cdot,\xi)$ need not be continuous at $t\in\gamma$. Given a flow map $f$, the upper-right Dini derivative of $V$ along $f$ at $(t,\xi,\mu)\in \R_{\ge 0}\times \R^n\times \R^m$ is
\begin{align*}
 D^{+}_{f}V(t,\xi,\mu):=\limsup_{h \to 0^+}\frac{V(t+h,\xi+h f(t,\xi,\mu))-V(t,\xi)}{h}
 \end{align*}
In our main result, we will suppose that the parametrized family $\{\Sigma_{\lambda}=(\gamma_{\lambda},f_{\lambda},g_{\lambda})\}_{\lambda \in \Lambda}$ and the functions $h^0,h \in \mathcal{H}$ satisfy the following assumption. 
\begin{as} 
  \label{ass:issV}
  There exists a parametrized family $\{V_{\lambda}\}_{\lambda\in \Lambda}$ of functions $V_{\lambda}\in \mathcal{V}(\gamma_{\lambda})$ such that 
  \begin{enumerate}[a)]
  \item there exist $\phi_1,\phi_2\in \Ki$ so that for all $\lambda\in \Lambda$, $t\ge 0$ and $\xi \in \R^n$,\label{item:Vphi12}
    \begin{align} 
      \label{eq:bound1}
      \phi_1(h(t,\xi))\le V_{\lambda}(t,\xi)\le \phi_2(h^o(t,\xi));
    \end{align}
  \item there exist $\chi\in\Ki$ and $\pi \in \Ki$ such that for each $\lambda \in \Lambda$ there exist a Carath\'eodory function $\varphi_{\lambda}:\R_{\ge 0}\times \R_{\ge 0}\to \R$ and a function $\alpha_{\lambda}:\R_{\ge 0}\times \R_{\ge 0} \to \R_{\ge 0}$ such that the following hold for all $t\ge 0$, $\xi \in \R^n$ and $\mu \in \R^m$: \label{item:Vimply}
   \begin{enumerate}[i)]   
   \item \label{item:lyapbound}
   $D^{+}_{f_{\lambda}}V_{\lambda}(t,\xi,\mu) \le -\varphi_{\lambda}(t,V_{\lambda}(t,\xi))$ if $t\notin\gamma_\lambda$ and $V_{\lambda}(t,\xi)\ge \chi(|\mu|)$;
   \item \label{item:lyapbound2}
      $V_{\lambda}(t,\xi+g_{\lambda}(t,\xi,\mu)) \le \alpha_{\lambda}(t, V_{\lambda}(t^-,\xi) )$ if $t\in\gamma_\lambda$ and $V_{\lambda}(t^-,\xi)\ge \chi(|\mu|)$;
      \item \label{item:lyapbound3}
    $V_{\lambda}(t,\xi+g_{\lambda}(t,\xi,\mu))\le \pi(|\mu|)$ if $t\in \gamma_{\lambda}$ and $V_{\lambda}(t^-,\xi) \le \chi(|\mu|).$
  \end{enumerate}
  \end{enumerate}
\end{as}
We note that no regularity condition is assumed on the functions $\alpha_{\lambda}$ appearing in~\ref{item:lyapbound2}).

For $\varphi_{\lambda}$, $\alpha_{\lambda}$ and $\gamma_{\lambda}\in\Gamma$ as in Assumption~\ref{ass:issV}, we consider the following parametrized family of one-dimensional differential/difference inclusion systems, which we henceforth call comparison systems,
\begin{subequations}
  \label{eq:isc}
  \begin{align}
    \label{eq:is-ctc}
    \dot{z}(t) &\in(-\infty,-\varphi_{\lambda}(t,z(t))], & 
    t\notin \gamma_{\lambda},    \displaybreak[0] \\
    \label{eq:is-stc}
    z(t) &\in[0,\alpha_{\lambda}(t,z(t^-))], & 
    t\in \gamma_{\lambda},
  \end{align}
\end{subequations}
We say that a function $z:I_z \to \R_{\ge 0}$, with $I_z=[t_0,T_z)$ is a solution of (\ref{eq:isc}) corresponding to $\lambda \in \Lambda$, initial time $t_0 \ge 0$ and initial condition $z_0\ge 0$ if i) $z(t_0) = z_0$, ii) if $J = [t_1,t_2) \subset I_z$ has no points of $\gamma_{\lambda}$ in its interior, then $z$ is locally absolutely continuous on $J$ and $\dot{z}(t)\le -\varphi_\lambda(t,z(t))$ for almost all $t\in J$, and iii) for every $t\in \gamma \cap (t_0,T_z)$, it happens that $z(t^-)$ exists and $0\le z(t) \le \alpha_\lambda(t,z(t^-))$. We will employ $\C_{\lambda}(t_0,z_0)$ to denote the set of solutions $z$ of \eqref{eq:isc} corresponding to $\lambda\in \Lambda$, initial time $t_0$ and initial condition $z_0$. For every $\lambda\in \Lambda$, $t_0\ge 0$ and $z_0\ge 0$, note that the definition of solution requires that for all $z\in\C_{\lambda}(t_0,z_0)$, $z(t) \ge 0$ for all $t\in I_z$.
We say that the parametrized family (\ref{eq:isc}) is weakly or strongly GUAS if there exists a function $\beta \in \KL$ such that every $z\in\C_{\lambda}(t_0,z_0)$ with $\lambda \in \Lambda$, $t_0\ge 0$ and $z_0\ge 0$ satisfies for all $t\in I_z$, respectively
\begin{align} 
  \label{eq:wbeta}
  z(t) &\le \beta\left (z_0,t-t_0\right) &\text{(weak)},\\
  \label{eq:sbeta}
  z(t) &\le \beta\left (z_0,t-t_0+n^{\gamma_{\lambda}}_{(t_0,t]}\right) &\text{(strong)}.
\end{align}
\begin{rem}
  \label{rem:equal} 
  If for some $\lambda\in \Lambda$, $\alpha_{\lambda}(t,r)$ is nondecreasing in $r$ for every $t \in \gamma_\lambda$ then each $z\in \mathcal{C}_{\lambda}(t_0,z_0)$ can be bounded from above by some solution $w$ of the impulsive system
\begin{subequations}
  \label{eq:isce}
  \begin{align}
    \label{eq:is-ctce}
    \dot{w}(t) &=-\varphi_{\lambda}(t,w(t)), & 
    t\notin \gamma_{\lambda},    \displaybreak[0] \\
    \label{eq:is-stce}
    w(t) &=\alpha_{\lambda}(t,w(t^-)), & 
    t\in \gamma_{\lambda},
  \end{align}
  \end{subequations}
  which satisfies $w(t_0)=z_0$. Then, when checking the GUAS of (\ref{eq:isc}), for those parameters $\lambda$ for which $\alpha_{\lambda}(t,r)$ is nondecreasing in $r$ it is sufficient to only consider the solutions of (\ref{eq:isce}) instead of all the solutions of (\ref{eq:isc}), i.e.  we can replace (\ref{eq:isc}) by (\ref{eq:isce}). The latter fact can be shown following the proof of Theorem~1.4.3 in \cite{lakbai_book89} replacing left-continuity by right-continuity and relaxing the continuity of the flow map to a Carath\'eodory condition. When $\alpha_{\lambda}(t,r)$ is not nondecreasing in $r$, such a simplification is not possible. \mer
  \end{rem}

Our main result is the following.
\begin{teo} 
  \label{thm:main} 
  Consider a parametrized family $\Sigma_{\Lambda}=\{\Sigma_{\lambda}=(\gamma_{\lambda},f_{\lambda},g_{\lambda})\}_{\lambda\in \Lambda}$ of impulsive systems with inputs. Let $h^o,h \in \mathcal{H}$ and let Assumption \ref{ass:issV} hold. Then $\Sigma_{\Lambda}$ is weakly or strongly $(h^o,h)$-ISS if the family (\ref{eq:isc}) of parametrized systems is, respectively, weakly or strongly GUAS.
\end{teo}
\begin{IEEEproof}
Assume that the parametrized family (\ref{eq:isc}) is weakly or strongly GUAS, and let $\beta \in \KL$ be as in (\ref{eq:wbeta})  or (\ref{eq:sbeta}), respectively. Let $b>0$, let $\lambda\in \Lambda$, let $t_0\ge 0$, let $x_0\in \R^n$, let $u\in\U$ be such that $\|u\|_{\gamma_{\lambda}}\le b$ and let $x \in \T_{\Sigma_\lambda}(t_0,x_0,u)$, $x : [t_0,T_x) \to \R^n$. 

Define the function $v_\lambda : [t_0, T_x) \to \R_{\ge 0}$ via $v_{\lambda}(t)=V_{\lambda}(t,x(t))$. From the facts that $V\in \mathcal{V}(\gamma_{\lambda})$ and $x$ is a solution of the impulsive system $\Sigma_{\lambda}$ it follows that $v_{\lambda}$ restricted to any interval $J = [t_1,t_2)\subset [t_0,T_x)$ without points of $\gamma_{\lambda}$ in its interior is locally absolutely continuous (note that this implies that $v_{\lambda}$ is right-continuous). It also follows that for each $\tau \in \gamma_{\lambda}\cap (t_0,T_x)$, then $v_{\lambda}(\tau^-)=\lim_{t\to\tau^-}V_{\lambda}(t,x(t)) = \lim_{t\to\tau^-} V_{\lambda}(t,x(\tau^-)) = V_{\lambda}(\tau^-,x(\tau^-))$. 

If $v_\lambda(t) \ge  \chi(b)$ for all $t\in [t_0,T_x)$, define $t_1 := T_x$. Otherwise, let $t_1 :=\inf\{t\in [t_0,T_x) : v_{\lambda}(t)< \chi(b)\}$. Then, $v_{\lambda}(t)\ge  \chi(b)$ for all $t\in [t_0,t_1)$. If $t_1 < T_x$, also $v_{\lambda}(t_1) \le \chi(b)$ by right-continuity. 

Due to Assumption \ref{ass:issV}\ref{item:lyapbound}) we have that
\begin{align}
\dot{v}_{\lambda}(t)\le -\varphi(t,v_{\lambda}(t))\;\text{for almost all} \;t\in [t_0,t_1)\setminus \gamma_{\lambda}.
\end{align}
Assumption \ref{ass:issV}\ref{item:lyapbound2}) implies that
\begin{align}
0\le v_{\lambda}(t)\le \alpha(t,v_{\lambda}(t^-)) \quad \forall t\in (t_0,t_1)\cap \gamma_{\lambda}.
\end{align}
In consequence, the function $v_\lambda$ restricted to $[t_0,t_1)$, which we still denote by $v_\lambda$, satisfies $v_\lambda \in \C_{\lambda}(t_0,v_\lambda(t_0))$ 
and then
\begin{align}\label{eq:wbound1}
v_{\lambda}(t)\le \beta\left(v_{\lambda}(t_0),t-t_0\right)\quad \forall t\in [t_0,t_1),
\end{align}
if (\ref{eq:isc}) is weakly GUAS and 
\begin{align}\label{eq:sbound1}
v_{\lambda}(t)\le \beta\left(v_{\lambda}(t_0),t-t_0+n^{\gamma_{\lambda}}_{(t_0,t]}\right)\quad \forall t\in [t_0,t_1),
\end{align}
if it is strongly GUAS. 

Next, consider the case $t_1 < T_x \le \infty$. 
Let $\nu=\max\{\pi,\chi\}\in \Ki$. We will prove that 
\begin{align} \label{eq:bound2}
v_{\lambda}(t)\le 2\beta(\nu(b)),0)\quad \forall t\in [t_1,T_x).
\end{align}
Note that (\ref{eq:bound2}) is valid at $t=t_1$ because we already know that $v_{\lambda}(t_1) \le \chi(b)$ and $\chi(b) \le \nu(b) \le \beta(\nu(b),0)$. For a contradiction, suppose that there exists $t^* \in (t_1,T_x)$ such that $v_{\lambda}(t^*)>2\beta(\nu(b)),0)$. Let $t_2=\inf\{t\in [t_1,T_x) : v_{\lambda}(t)>2\beta(\nu(b),0) \}$. Since $v_{\lambda}$ is right-continuous, $v_{\lambda}(t_2)\ge 2\beta(\nu(b)),0)>\chi(b)$, and then $t_1<t_2$. Define $\hat t_1=\sup\{t\in[t_1,t_2]:v_{\lambda}(t)\le \chi(b)\}$. If $v_{\lambda}(\hat t_1)>\chi(b)$, then $\hat t_1>t_1$ [since $v_{\lambda}(t_1)\le \chi(b)$], $\hat t_1 \in \gamma_{\lambda}$ and $v_{\lambda}(\hat t_1^-)\le \chi(b)$. Then $v_{\lambda}(\hat t_1)\le \pi(b)$ due to Assumption \ref{ass:issV}\ref{item:lyapbound3}).  The latter implies that $v_{\lambda}(\hat t_1)\le \nu(b)$. If $v_{\lambda}(\hat t_1)\le \chi(b)$, then $\hat t_1<t_2$ and, due to right-continuity of $v_{\lambda}$, $v_{\lambda}(\hat t_1)=\chi(b)\le \nu(b)$. In both cases we have that $v_{\lambda}(t)\ge \chi(b)$ for all $t\in [\hat t_1,t_2]$, $v_{\lambda}(\hat t_1)\le \nu(b)$ and $v_{\lambda}(t_2)\ge 2\beta(\nu(b),0)$. By right-continuity, there is $t_2<\hat{t_2}<T_x$ such that  $v_{\lambda}(t)\ge \chi(b)$ for all $t\in [\hat t_1,\hat t_2)$. Reasoning as in the beginning of the proof, it follows that for all $t\in [\hat t_1,\hat t_2)$, $v_{\lambda}(t)\le \beta(\nu(b),t-\hat t_1)$ when the stability of (\ref{eq:isc}) is weak and $v_{\lambda}(t)\le \beta(\nu(b),t-\hat t_1+n^{\gamma_{\lambda}}_{(\hat t_1,t]})$ when it is strong. In particular,  $v_{\lambda}(t_2)\le \beta(\nu(b),0)<2\beta(\nu(b),0)$. Since we have arrived to a contradiction, then (\ref{eq:bound2}) holds. 
Define $\tilde \rho(r)=2\beta(\nu(r),0)$ for all $r\ge 0$, then $\tilde \rho\in \Ki$. From (\ref{eq:wbound1}) and (\ref{eq:bound2}) it follows that
\begin{align*}
v_{\lambda}(t)\le \max\left \{\beta\left(v_{\lambda}(t_0),t-t_0\right), \tilde \rho(b) \right \}\quad \forall  t\in [t_0,T_x),
\end{align*}
in the weak case, and from From (\ref{eq:sbound1}) and (\ref{eq:bound2}) we have that
\begin{align*}
v_{\lambda}(t)\le \max\left \{\beta\left(v_{\lambda}(t_0),t-t_0+n^{\gamma_{\lambda}}_{(t_0,t]}\right), \tilde \rho(b) \right \}
\end{align*}
for all $t\in [t_0,T_x)$ in the strong one.

Then, from Assumption~\ref{ass:issV}\ref{item:Vphi12}) and defining $\hat \beta(r,t)=\phi_1^{-1}(\beta(\phi_2(r),t))$ and $\rho=\phi_{1}^{-1}\comp \tilde \rho$, it follows that for all $t\in [t_0,T_x)$:
\begin{align} \label{eq:wmaxISS}
h(t,x(t))\le \max\left \{\hat \beta\left(h^o(t_0,x(t_0)),t-t_0\right), \rho(b)\right \}
\end{align}
in the weak case, and
\begin{multline} \label{eq:smaxISS}
h(t,x(t)) \\\le \max\left \{\hat \beta\left(h^o(t_0,x(t_0)),t-t_0+n^{\gamma_{\lambda}}_{(t_0,t]}\right), \rho(b)\right \}
\end{multline}
in the strong case.  Since (\ref{eq:wmaxISS}) and (\ref{eq:smaxISS}) hold for any positive constant $b$ such that $\|u\|_{\gamma_{\lambda}}\le b$, we arrive to
\begin{align*}
h(t,x(t))&\le \max\left \{\hat \beta\left(h^o(t_0,x(t_0)),t-t_0\right), \rho(\|u\|_{\gamma_{\lambda}})\right \}\\
&\le \hat \beta\left(h^o(t_0,x(t_0)),t-t_0 \right)+ \rho(\|u\|_{\gamma_{\lambda}}),
\end{align*}
when the parametrized family (\ref{eq:isc}) is weakly GUAS, and to
\begin{align*}
\lefteqn{h(t,x(t))} \hspace{2mm}\\
&\le \max\left \{\hat \beta\left(h^o(t_0,x(t_0)),t-t_0+n^{\gamma_{\lambda}}_{(t_0,t]}\right), \rho(\|u\|_{\gamma_{\lambda}})\right \}\\
&\le \hat \beta\left(h^o(t_0,x(t_0)),t-t_0+n^{\gamma_{\lambda}}_{(t_0,t]}\right)+ \rho(\|u\|_{\gamma_{\lambda}}),
\end{align*}
when the parametrized family (\ref{eq:isc}) is strongly GUAS.
\end{IEEEproof}
Theorem~\ref{thm:main} shows how assessing $(h^o,h)$-ISS of a family of impulsive systems can be performed by reducing the problem to assessing the GUAS of the family of comparison systems (\ref{eq:isc}). The latter systems have the advantage of being scalar and without inputs. In the next section, we employ Theorem~\ref{thm:main} in order to generalize several existing results. Comparison-type results have been recently employed in \cite{aichen_nahs17} for impulsive systems with no inputs and in \cite{liuliu_scl12} for impulsive (switched) systems with inputs. In both papers, the comparison systems are given by impulsive differential equations like (\ref{eq:isce}) instead of by differential/difference inclusions as (\ref{eq:isc}). Refs.~\cite{liuliu_scl12} and~\cite{aichen_nahs17} require that the functions playing the role of our $\alpha_{\lambda}(t,r)$ in Assumption~\ref{ass:issV} be nondecreasing in $r$. This requirement allows the use of the simpler comparison systems (see Remark~\ref{rem:equal}). Assuming that the functions $\alpha_{\lambda}(t,r)$ are nondecreasing in $r$ seems a nonnatural and restrictive assumption in a nonlinear context (see Theorems~\ref{thm:sciss1} and~\ref{thm:sw}).

\begin{rem}
  \label{rem:further}
  The type of decaying term in the ISS property considered will correspond with the type of decaying term that bounds the trajectories of the comparison systems (\ref{eq:isc}). Ideas analogous to those employed in the proof of Theorem~\ref{thm:main} could also be employed to derive results when the decaying term converges in finite-time \cite{liho_auto19} or is exponential \cite{naghes_scl08}. Moreover, the two-measure framework can be employed when stability is only practical \cite{ghanmi_mmas16}.\mer
\end{rem}
\begin{example}
  \label{ex:family}
  Let $\eta\in\Ki$ satisfy $\eta(r) \le r$ for $r\ge 1$. Consider the parametrized family of scalar impulsive systems with inputs $\Sigma_{\Lambda} = \{\Sigma_\lambda = (\gamma_\lambda,f_\lambda,g_\lambda)\}_{\lambda \in \Lambda}$, with $\Lambda = (1,\infty)$, 
  \begin{align*}
    \gamma_\lambda = \{\tau_k^\lambda &\}_{k=1}^\infty, \quad \tau_0^\lambda = 0, \quad \tau_{k}^\lambda = \tau_{k-1}^\lambda + \frac{1}{\lambda k},\\
    f_\lambda(t,\xi, &\mu) = -2 |\xi|^{\lambda}\, \sign(\xi) + \mu^\lambda,\\
    g_\lambda(t,\xi, &\mu) =
      \begin{cases}
        -\xi + \frac{\sign(\xi)
          p_\lambda(|\xi|,k)}{\exp(1/(\lambda
          k))} &\text{for }t=\tau_k^\lambda,\\
        -\xi &\text{otherwise,}
      \end{cases}
\\
    p_\lambda(r,k) &=
    \begin{cases}
      \min\{\eta(r),q_\lambda(r,k)\} &\text{if }r < \left[ \frac{\lambda k}{\lambda -1} \right]^{\frac{1}{\lambda - 1}},\\ 
      r  &\text{otherwise,}
    \end{cases}\\
    q_\lambda(r,k) &= \left[ r^{1-\lambda} - \dfrac{\lambda - 1}{\lambda k} \right]^{\frac{1}{1-\lambda}},
  \end{align*}
  Note that for every $\lambda \in (1,\infty)$ and $k\in\N$, $q_\lambda(\cdot,k) : \left(0, \left[\frac{\lambda k}{\lambda -1}\right]^{\frac{1}{\lambda-1}} \right) \to \R_{>0}$ is continuous, increasing and satisfies $\lim_{r\to 0^+} q_\lambda(r,k) = 0$. Consider $V(\xi) = |\xi|$ and let $h^o(t,\xi) = h(t,\xi) = |\xi|$. The family $\{V_\lambda\}_{\lambda\in\Lambda}$ of functions defined via $V_\lambda(t,\xi) = V(\xi)$ for all $\lambda\in\Lambda$ satisfies Assumption~\ref{ass:issV} with $\phi_1 = \phi_2 = \chi = \id$, $\pi = \max\{\eta,\id\}$, $\varphi_\lambda(t,r) = r^\lambda$, $\alpha_\lambda(t,r) = r+g_\lambda(t,r,0)$. Moreover, $\alpha_\lambda(t,r)$ is nondecreasing in $r$ for every fixed $t\ge 0$ and $\lambda \in \Lambda$, but may be discontinuous at $r = \left[ \frac{\lambda k}{\lambda -1} \right]^{\frac{1}{\lambda - 1}}$ when $t=\tau_k^\lambda$. For each $\lambda\in\Lambda$, $t_0 \ge 0$ and $w_0 = w(t_0) \ge 0$, the scalar impulsive system (\ref{eq:isce}) has a unique forward-in-time solution. Let $\ell \in \N$ satisfy $\tau_{\ell-1}^\lambda \le t_0 < \tau_{\ell}^\lambda$, then the solution is given by:
  \begin{itemize}
  \item If $t \in [t_0, \tau_{\ell}^\lambda)$ and $w(t_0) > 0$, then\\ $w(t) = \left[ w(t_0)^{1-\lambda} + (\lambda -1)(t-t_0) \right]^{\frac{1}{1-\lambda}}$,
  \item If $t \in [\tau_k^\lambda, \tau_{k+1}^\lambda)$, $k\ge\ell$ and
    $w(t_0) > 0$, then\\
      $w(t) = \left[ w(\tau_k^\lambda)^{1-\lambda} + (\lambda -1)(t-\tau_k^\lambda) \right]^{\frac{1}{1-\lambda}},$
  \item If $t = \tau_k^\lambda \ge \tau_{\ell}^\lambda$, 
    then $w(t) = p_\lambda(w(t^-),k) e^{-\frac{1}{\lambda k}}$.
  \item Otherwise, $w(t) = 0$. 
  \end{itemize}
  The solutions of (\ref{eq:isce}) have the following property: whenever the system has undergone flow for a whole interval $[\tau_k^\lambda,\tau_{k+1}^\lambda)$, the jump occurring at $t=\tau_{k+1}^\lambda$ reverts the value of the state to that at the beginning of this flow interval or to a smaller one, and applies an exponential decrease to the resulting value. From these solution equations, it follows that if $0 \le \tau_{\ell-1}^\lambda \le t_0 < \tau_\ell^\lambda < \tau_k^\lambda < \tau_{k+1}^\lambda$, then
  \begin{align*}
    w(\tau_{k+1}^\lambda) &= p_\lambda(w((\tau_{k+1}^\lambda)^-),k+1) e^{-\frac{1}{\lambda (k+1)}}\\
    &\le w(\tau_k^\lambda) e^{-\frac{1}{\blambda(k+1)}} = w(\tau_k^\lambda) e^{-(\tau_{k+1}^\lambda - \tau_k^\lambda)}\\
    &\le w(\tau_\ell^\lambda) e^{-(\tau_{k+1}^\lambda - \tau_\ell^\lambda)} \le \pi(w(t_0)) e^1 e^{-(\tau_{k+1}^\lambda - t_0)},
  \end{align*}
  where we have used the fact that $\tau_\ell^\lambda - t_0 \le \tau_\ell^\lambda - \tau_{\ell-1}^\lambda = \frac{1}{\lambda \ell} \le 1$. Since $w(\cdot)$ is nonincreasing over each interval $[\tau_k^\lambda, \tau_{k+1}^\lambda)$, we finally obtain that, for every $\lambda \in \Lambda$,
  \begin{align}
    \label{eq:wexpbnd}
    w(t) &\le \pi(w(t_0)) e^2 e^{-(t-t_0)}\quad \text{for all }t\ge t_0.
  \end{align}
  This shows that the family (\ref{eq:isce}) is GUAS. In view of Remark~\ref{rem:equal} and Theorem~\ref{thm:main}, it follows that $\Sigma_\Lambda$ is weakly ISS. Note that $\Gamma_\Lambda = \{\gamma_\lambda\}_{\lambda \in \Lambda}$ is not UIB and that we cannot conclude that (\ref{eq:isce}) is strongly GUAS.
\end{example}

We note that no existing ISS criteria can be applied to the family of systems considered in Example~\ref{ex:family} since, to the best of our knowledge, all existing criteria require fixed flow and jump maps.

\section{Sufficient Conditions for ISS}
\label{sec:suff-cond-iss}

In this section we derive criteria for the weak or strong $(h^o,h)$-ISS of parametrized families of impulsive systems under different hypotheses on the functions $\varphi_\lambda$ and $\alpha_\lambda$ appearing in (\ref{eq:isc}) and on the family of impulse-time sequences $\Gamma_{\Lambda}=\{\gamma_{\lambda}:\lambda\in \Lambda\}$. We also show how each of the results presented generalizes existing results for impulsive as well as some for nonimpulsive systems.

We begin with a result for cases when the continuous part of the system is stabilizing. Given a parametrized family of differential equations $\dot{w}=-\varphi_{\lambda}(t,w)$, $\lambda \in \Lambda$, with $\varphi_{\lambda}:\R_{\ge 0}\times \R_{\ge 0}\to \R$ a Carath\'eodory function and such that $\varphi_{\lambda}(t,0)=0$ for all $t\ge 0$, we say that this family is GUAS if there exists $\beta \in \KL$ such that for every $\lambda$, each maximally defined forward-in-time solution $w$ of $\dot{w}(t)=-\varphi_{\lambda}(t,w(t))$ is defined for all $t\ge t_0$, where $t_0$ is its initial time, and satisfies $w(t)\le \beta(w(t_0),t-t_0)$ for all $t\ge t_0$.
\begin{teo}
  \label{teo:arbitrary}
 Let $h^o,h\in \mathcal{H}$. Let $\Sigma_{\Lambda} = \{\Sigma_{\lambda} = (\gamma_{\lambda}, f_{\lambda}, g_{\lambda}) \}_{\lambda\in \Lambda}$ satisfy Assumption~\ref{ass:issV} with $\alpha_{\lambda}(t,r)=r$ for all $t,r\ge 0$ and $\lambda \in \Lambda$. Suppose that the family of differential equations $\dot{w}=-\varphi_{\lambda}(t,w)$ is GUAS. Then $\Sigma_{\Lambda}$ is weakly $(h^0,h)$-ISS. If, in addition, $\Gamma_{\Lambda} = \{\gamma_\lambda : \lambda \in \Lambda\}$ is UIB, then  $\Sigma_{\Lambda}$ is strongly $(h^0,h)$-ISS.
\end{teo}
\begin{IEEEproof} From Remark \ref{rem:equal} it follows that for checking that  (\ref{eq:isc}) is weakly  GUAS, it is sufficient to consider only the solutions of the family of impulsive systems (\ref{eq:isce}) with $\alpha_{\lambda}(t,r)=r$. Since the solutions of the latter family of impulsive systems coincide with those of the GUAS family of differential equations $\dot{w}=-\varphi_{\lambda}(t,w)$,  the weak GUAS of (\ref{eq:isc}) follows, and therefore $\Sigma_{\Lambda}$ is weakly $(h^o,h)$-ISS due to Theorem \ref{thm:main}. 

That $\Sigma_{\Lambda}$ is strongly $(h^o,h)$-ISS when $\Gamma_{\Lambda}$ is UIB follows from Proposition \ref{prop:equivalence}.
\end{IEEEproof}
Theorem \ref{teo:arbitrary}) contains as a particular case the well-known sufficient conditions for the ISS of nonimpulsive time-invariant systems given in terms of ISS Lyapunov functions (Lemma 2.14 in \cite{sonwan_scl95}). In fact, if $V$ is an ISS Lyapunov function (as per Definition 2.2 in \cite{sonwan_scl95}) for the system $\dot{x}=f(x,u)$, then the single system $\Sigma=(\gamma,f,g)$ (we drop the parameter $\lambda$ here) with $\gamma$ any impulse-time sequence and $g\equiv 0$, satisfies the hypotheses of Theorem \ref{teo:arbitrary} with $\varphi(t,\xi)\equiv \bar \varphi(\xi)$, where $\bar \varphi$ is a continuous positive definite function. Since $\dot w=-\bar \varphi(w)$ is GUAS, then $\Sigma=(\gamma,f,g)$ is weakly ISS and therefore $\dot{x}=f(x,u)$ is ISS. 

Theorem~\ref{teo:arbitrary} also contains as a particular case Theorem~2 of \cite{cheyan_tac17}, which in turn generalizes Theorem~1 of \cite{ninghe_scl12}. These results give sufficient conditions for ISS of the nonimpulsive systems $\dot{x}=f(t,x,u)$ based on a Lyapunov function with an indefinite derivative. Specifically, if we consider the family of systems which consists of a single system whose impulsive part is $g(t,\xi,\mu) \equiv 0$ and take $h^0(t,\xi)=h(t,\xi)=|\xi|$, the function $V$ appearing in \cite[Theorem~2]{cheyan_tac17} satisfies Assumption \ref{ass:issV} with  $\varphi(t,r) = -g(t) r$, $\alpha(t,r) = r$, $\chi=\rho$ and $\pi=\alpha_2\comp \alpha_1^{-1}\comp \rho$, where the latter $g$, $\rho$, $\alpha_1$, and $\alpha_2$ are the functions appearing in \cite[Theorem~2]{cheyan_tac17}. The ISS of $\dot{x}=f(t,x,u)$ follows from Theorem~\ref{teo:arbitrary} since the conditions imposed on $g$ in items 3)--4) of \cite[Theorem~2]{cheyan_tac17} imply that the system $\dot w = g(t) w = -\varphi(t,w)$ is GUAS.


The following criterion involves an average dwell-time condition on the impulse-time sequences when the continuous-time dynamics is $(h^0,h)$-ISS and a reverse average dwell-time one when the discrete-time dynamics is stabilizing. Given $N_0\in \N$ and $\tau_D>0$, $\Gamma_{\textsc{adt}}[N_0,\tau_D]$ (resp. $\Gamma_{\textsc{radt}}[N_0,\tau_D]$) is the set of impulse-time sequences $\gamma$ such that $n^{\gamma}_{(s,t]}\le N_0+\frac{t-s}{\tau_D}$ (resp. $n^{\gamma}_{(s,t]}\ge \frac{t-s}{\tau_D}-N_0$) for all $0\le s<t$. Note that $\Gamma_{\textsc{adt}}[N_0,\tau_D]$ is UIB. From this point on, we employ the following conventions: $0\ln(0)=0$ and $r \ln(0)=-\infty$ if $r>0$.
\begin{teo}
  \label{thm:avdt-radt}
  Let $\Sigma_{\Lambda} = \{\Sigma_{\lambda} = (\gamma_{\lambda}, f_{\lambda}, g_{\lambda}) \}_{\lambda\in \Lambda}$ and $h^o, h\in \mathcal{H}$ satisfy Assumption~\ref{ass:issV} with $\varphi_{\lambda}(t,r)=\phi_{\lambda}(t)r$, $\phi_{\lambda}$ Lebesgue measurable and locally integrable, and $\alpha_{\lambda}(t,r)=d\cdot r$ with $d\ge 0$. 
  Suppose that there exist $\kappa>0$ and $c\in \R$ such that for all $\lambda\in \Lambda$ the solutions of the linear equation $\dot{w}=-\phi_{\lambda}(t)w$ satisfy
  \begin{align} \label{eq:bound3}
    |w(t)|\le \kappa e^{-c (t-t_0)}|w(t_0)|\quad \forall t\ge t_0\ge 0.
  \end{align}
  Then, $\Sigma_{\Lambda}$ is 
  \begin{enumerate}[a)]
  \item \label{item:wISS-adt}weakly $(h^o,h)$-ISS if there exist $\eta,\mu > 0$ such that the following condition holds for every $\gamma \in \Gamma_\Lambda$:
    \begin{align}
      \label{eq:hespanha}
      n^{\gamma}_{(t_0,t]} \ln(d) - (c-\eta)(t-t_0) &\le \mu, \quad \forall t\ge t_0 \ge 0;
    \end{align}
  \item \label{item:sISS-adt}strongly $(h^o,h)$-ISS if there exist $\eta,\mu > 0$ such that the following condition holds for every $\gamma \in \Gamma_\Lambda$:
    \begin{align}
      \label{eq:shespanha}
      n^{\gamma}_{(t_0,t]} [\ln(d)+\eta] - (c-\eta)(t-t_0) &\le \mu, \quad \forall t\ge t_0 \ge 0.
    \end{align}
  \end{enumerate}
\end{teo}
\begin{IEEEproof} 
  Since $\alpha_{\lambda}(t,r)$ is nondecreasing in $r$ for every $\lambda$, from Remark \ref{rem:equal} we have that for establishing the weak or strong GUAS of the comparison systems (\ref{eq:isc}) it is sufficient to establish that of the systems (\ref{eq:isce}) with $\varphi_{\lambda}(t,r)=-\phi_{\lambda}(t) r$ and $\alpha_{\lambda}(t,r)=d\cdot r$. Let $\lambda \in \Lambda$, $t_0\ge 0$, $z_0 \ge 0$ and let $z:I_z\to \R_{\ge 0}$, $I_z=[t_0,T_z)$, be a solution of (\ref{eq:isce}) with $z(t_0)=z_0$. By solving (\ref{eq:isce}) it easily follows that
  \begin{align*}  
    z(t)= z_0 e^{-\int_{t_0}^t\phi_{\lambda}(s)\:ds+n^{\gamma_{\lambda}}_{(t_0,t]}\ln(d)}\quad \forall t\in I_z\cap[t_0,\infty).
  \end{align*}
  Since $w(t)=z_0 e^{-\int_{t_0}^t\phi_{\lambda}(s)\:ds}$ is the solution of the initial value problem $\dot{w}=-\phi_{\lambda}(t)w$, $w(t_0)=z_0$ and taking into account (\ref{eq:bound3}), we then have that 
  \begin{align} 
    \label{eq:bound4}
    z(t)\le \kappa z_0 e^{-c(t-t_0)+n^{\gamma_{\lambda}}_{(t_0,t]}\ln(d)}\quad \forall t\in I_z\cap[t_0,\infty).
  \end{align}

  \ref{item:wISS-adt}) Employing~(\ref{eq:hespanha}), from (\ref{eq:bound4}) it follows that
  \begin{align*}
    z(t)\le \kappa e^{\mu} z_0 e^{-\eta(t-t_0)}\quad \forall t\in I_z\cap[t_0,\infty).
  \end{align*}
  This shows that (\ref{eq:isce}) is weakly GUAS and consequently the same holds for (\ref{eq:isc}). By Theorem~\ref{thm:main}, then $\Sigma_\Lambda$ is weakly $(h^o,h)$-ISS.

  \ref{item:sISS-adt}) Employing~(\ref{eq:shespanha}), from~(\ref{eq:bound4}) it follows that
  \begin{align*}
    z(t)\le \kappa e^{\mu} z_0 e^{-\eta(t-t_0+n^{\gamma_{\lambda}}_{(t_0,t]})}\quad \forall t\in I_z\cap[t_0,\infty).    
  \end{align*}
  This shows that (\ref{eq:isce}) is strongly GUAS and therefore (\ref{eq:isc}) is so. By Theorem~\ref{thm:main}, then $\Sigma_\Lambda$ is strongly $(h^o,h)$-ISS.
\end{IEEEproof}
We note that Theorem~1 of \cite{heslib_auto08} follows from \ref{item:wISS-adt}) of Theorem~\ref{thm:avdt-radt}, by setting $h^o(t,\xi)=h(t,\xi)=|\xi|$ and $\phi(t)\equiv c$, with $c\in \R$.

From the proof of Theorem~\ref{thm:avdt-radt} it easily follows that if instead of (\ref{eq:bound3}) and (\ref{eq:hespanha}) we assume that there exist $\eta,\mu>0$ such that for every $\lambda \in \Lambda$ the following condition holds:
\begin{align}\label{ning}
n^{\gamma_{\lambda}}_{(t_0,t]} \ln(d) - \int_{t_0}^{t}\phi_{\lambda}(s)ds &\le \mu-\eta(t-t_0), \quad \forall t\ge t_0 \ge 0,
    \end{align}
    then $\Sigma_{\Lambda}$ is weakly $(h^o,h)$-ISS. Theorem 1 in \cite{ninhe_is18} follows from this observation, since the conditions therein imply the satisfaction of (\ref{ning}). 
\begin{cor}
  \label{cor:adt-radt}
  Under the assumptions of Theorem~\ref{thm:avdt-radt}, the family $\Sigma_\Lambda$ is strongly $(h^o,h)$-ISS if one of the following conditions holds:
  \begin{enumerate}[a)] 
  \item \label{item:avdt} $c>0$ and $\Gamma_{\Lambda}\subset \Gamma_{\textsc{adt}}[N_0,\tau_D]$ for some $N_0\in \N$ and some\footnote{if $d=0$, then any $\tau_D > 0$ is valid.} $\tau_D> \max\{\ln(d)/c,0\}$.
  \item \label{item:ravdt} $c\le 0$, $d<1$ and $\Gamma_{\Lambda}\subset \Gamma_{\textsc{radt}}[N_0,\tau_D]$ for some $N_0\in \N$ and some\footnote{if $cd=0$, then any $\tau_D > 0$ is valid.} $0<\tau_D<|\ln(d)/c|$.
  \end{enumerate}
\end{cor}
\begin{IEEEproof}
    Consider item~\ref{item:avdt}). Let $0 < \eta := \ln(d)/\tau_D < c$ and $\mu = N_0 \ln(d)$ if $d>1$, and $0 < \eta < c$ and $\mu > 0$ but otherwise both arbitrary if $0 \le d \le 1$. Then   
  $n^{\gamma}_{(t_0,t]}\le N_0+\frac{t-t_0}{\tau_D}$, and therefore, 
  \begin{align*}
    -c(t-t_0)+n^{\gamma}_{(t_0,t]}\ln(d) &\le \mu -\eta (t-t_0)
  \end{align*}
  in either case. This shows that (\ref{eq:hespanha}) is satisfied and by Theorem~\ref{thm:avdt-radt}, then $\Sigma_\Lambda$ is weakly $(h^o,h)$-ISS. Since $\Gamma_{\Lambda} \subset \Gamma_{\textsc{adt}}[N_0,\tau_D]$ is UIB, then $\Sigma_{\Lambda}$ is also strongly $(h^o,h)$-ISS by Proposition~\ref{prop:equivalence}.
  
  Next, consider item~\ref{item:ravdt}). In this case, $n^{\gamma}_{(t_0,t]}\ge \frac{t-t_0}{\tau_D}-N_0$ for all $0 \le t_0<t$ or, equivalently, $t-t_0 \le \tau_D N_0+ \tau_D n^{\gamma}_{(t_0,t]}$. Set $\hat d = |\ln(d)| - |c|\tau_D > 0$, $\bar\tau_D = \max\{\tau_d,1\}$, $\eta = \hat d/(2\bar\tau_D) > 0$ and $\mu = |c|\tau_D N_0 + \hat d/(2N_0) > 0$. Therefore,
  \begin{align}
    \label{eq:radt1}
    -c(t-t_0)+n^{\gamma}_{(t_0,t]} \ln(d) &\le -\hat d n^{\gamma}_{(t_0,t]} + |c| \tau_D N_0.
  \end{align}
  %
  We also have
  \begin{align*}
    \frac{\hat{d}}{2} n^{\gamma}_{(t_0,t]} &\ge \frac{\hat{d}}{2\bar\tau_D} n^{\gamma}_{(t_0,t]},\quad\text{and}\\
    \frac{\hat{d}}{2} n^{\gamma}_{(t_0,t]} &\ge \frac{\hat{d}}{2}\left[\frac{t-t_0}{\tau_D} - N_0\right] \ge \frac{\hat{d}}{2}\left[\frac{t-t_0}{\bar\tau_D} - N_0\right],
  \end{align*}
  and then
  \begin{align}
    \label{eq:radt2}
    \eta\left(t-t_0+n^{\gamma}_{(t_0,t]} \right) &\le \hat d n^{\gamma}_{(t_0,t]} + \frac{\hat d}{2N_0}
  \end{align}
  Adding up (\ref{eq:radt1}) and (\ref{eq:radt2}), it follows that (\ref{eq:shespanha}) is satisfied. By Theorem~\ref{thm:avdt-radt}, the result follows.
%
\end{IEEEproof}

Corollary~\ref{cor:adt-radt} generalizes Corollary~1 of \cite{heslib_auto08} and contains as particular case Theorem 1 in \cite{lili_mcs19}. In addition, we show that the same dwell-time conditions as in \cite{heslib_auto08} and \cite{lili_mcs19} actually ensure not only weak $(h^o,h)$-ISS but also strong $(h^o,h)$-ISS. This is perhaps not surprising for the average dwell-time condition given that $\Gamma_{\textsc{adt}}[N_0,\tau_D]$ is UIB; however, also strong $(h^o,h)$-ISS is established for the reverse average dwell-time condition.

For a given constant $\theta>0$, let $\Gamma_{\theta}$ and $\Gamma^{\theta}$ denote the classes of impulse-time sequences having a minimum and maximum dwell time given by $\theta$, respectively. More precisely, $\Gamma_\theta$ and $\Gamma^\theta$ are the classes of impulse-time sequences $\gamma=\{\tau_k\}_{k=1}^{N}$ which verify  $\tau_{k+1}-\tau_k\ge \theta$ for all $0\le k< N$, and $N=\infty$ and $\tau_{k+1}-\tau_k\le \theta$ for all $k\ge 0$, respectively. The following criterion involves a genuinely nonlinear condition on the Lyapunov-like function from Assumption~\ref{ass:issV}. 
\begin{teo} 
  \label{thm:sciss1} 
  Let $\Sigma_{\Lambda} = \{\Sigma_{\lambda} = (\gamma_{\lambda}, f_{\lambda}, g_{\lambda}) \}_{\lambda\in \Lambda}$ and $h^o, h\in \mathcal{H}$ satisfy Assumption~\ref{ass:issV} with $\varphi_\lambda(t,r)=\phi(t) \bar\varphi(r)$ for all $\lambda\in\Lambda$, $\phi : \R_{\ge 0}\to \R_{\ge 0}$ locally integrable and $\bar\varphi$ continuous, and $\alpha_\lambda(t,r) = \bar\alpha(r)$ for all $\lambda \in \Lambda$, $\bar\alpha$ continuous and positive definite. Then $\Sigma_{\Lambda}$ is strongly $(h^o,h)$-ISS if one of the following two conditions holds:
  \begin{enumerate}[a)]
  \item \label{item:phipd} $\bar\varphi$ is positive definite and there exists a constant $\theta>0$ such that $\gamma_{\lambda}\in \Gamma_{\theta}$ for all $\lambda \in \Lambda$ and
    \begin{align}\label{eq:integralb}
     \sup_{a>0} \int_a^{\bar \alpha(a)}\frac{ds}{\bar \varphi(s)}< \inf_{t\ge 0}\int_{t}^{t+\theta}\phi(s)\:ds=:M
    \end{align}
    with $M>0$.
  \item \label{item:phind} $-\bar\varphi$ is positive definite and there exists a constant $\theta>0$ such that for all $\lambda \in \Lambda$, $\gamma_{\lambda}\in \Gamma^{\theta}$ and
    \begin{align}
      \int_1^\infty \frac{ds}{-\bar\varphi(s)} &= \infty,\quad\text{and} \label{eq:integralr}\\ \inf_{a>0} \int_{\bar \alpha(a)}^a\frac{ds}{- \bar\varphi(s)} &> \sup_{t\ge 0}\int_{t}^{t+\theta}\phi(s)\:ds. \label{eq:integralr2}
    \end{align}
  \end{enumerate}
\end{teo}
The proof of Theorem~\ref{thm:sciss1} is given in Section~\ref{sec:pf-thm-sciss1}. Theorem~\ref{thm:sciss1} generalizes Theorems~1 and~3 in \cite{dasmir_siamjco13} and also gives far stronger conclusions. In fact, the ISS property ensured by Theorem~\ref{thm:sciss1} is stronger than that considered in \cite{dasmir_siamjco13} and, in addition, it is uniform with respect to both, initial time and the family of systems, while none of these uniformities is guaranteed in \cite{dasmir_siamjco13}, as explained in Remark~2 on p.~1970 therein. Very recently, results in \cite{fekbaj_ecc19} ensure a nonuniform version of weak ISS under condition a) of Theorem \ref{thm:sciss1} with $\phi(s)\equiv 1$ and under average dwell time between impulses. The latter average dwell-time condition is more general than the dwell-time condition of our results. However, the type of weak ISS considered therein is uniform neither in the initial time nor over the class of sequences.

\begin{example}
  Consider the family of scalar impulsive systems with a single input $\{\Sigma_{\gamma}=(\gamma,f,g)\}_{\gamma \in \Gamma^*}$, where $\Gamma^*$ is a class of impulse-time sequences,
  \begin{align*}
    f(t,\xi,\mu) &= \ell(|\xi|) (\mu + \xi),\\
    g(t,\xi,\mu) &= \delta(|\xi|) (\mu^3 + \xi) - \xi,
  \end{align*}
$\ell,\delta : \R_{\ge 0} \to \R_{\ge 0}$, $\ell$ is continuous, satisfies $\ell(r) = r/2$ if $0 \le r\le 1$, $1/2 \le \ell(r) \le r/2$ for $r\ge 1$, and
  \begin{gather*}
    \delta(r) :=
                \begin{cases}
                  r^2/4 -r/2 + 1/2 &\text{if }0 \le r <1,\\
                  1/(4r) &\text{if }r\ge 1.
                \end{cases}
  \end{gather*}
  The function $\delta$ is thus strictly decreasing and $\delta(0) = 1/2$. 
  We would like to determine conditions on $\Gamma^*$ and on the function $\ell$ so that the family of impulsive systems is strongly ISS. Therefore, we consider $h^0(t,\xi) = h(t,\xi) = |\xi|$. Define $\bar V(\xi) := |\xi|$ and $V_\gamma(t,\xi) \equiv \bar V(\xi)$. Then (\ref{eq:bound1}) is satisfied with $\phi_1 = \phi_2 = \id$. We have, for all $\xi,\mu \in \R$,
  \begin{align*}
    D_f^+ \bar V(t,\xi,\mu) &\le \ell(|\xi|) (|\mu| + |\xi|),\\
    \bar V(\xi + g(t,\xi,\mu)) &\le \delta(|\xi|) (|\mu|^3 + |\xi|).
  \end{align*}
  Define $\chi\in\Ki$ via $\chi(r) = r^3 + r$. If $|\xi| \ge \chi(|\mu|)$, then
  \begin{align*}
    D_f^+ \bar V(t,\xi,\mu) &\le 2\ell(|\xi|)|\xi|,\quad\text{and}\\
    \bar V(\xi + g(t,\xi,\mu)) &\le 2\delta(|\xi|) |\xi| \le 2\delta(0) |\xi| = |\xi|.
  \end{align*}
  If $|\xi| \le \chi(|\mu|)$, then
  \begin{align*}
    \bar V(\xi + g(t,\xi,\mu)) &\le \delta(0) (|\mu|^3 + \chi(|\mu|)) =: \pi(|\mu|),
  \end{align*}
  with $\pi\in\Ki$. Define $\bar\varphi(r) := -2\ell(r)r$ and $\bar\alpha(r) := 2\delta(r)r \le r$. Then, Assumption~\ref{ass:issV} is satisfied with $\varphi_\gamma(t,r) \equiv \bar\varphi(r)$ and $\alpha_\gamma(t,r) \equiv \bar\alpha(r)$. Note that during flows, the magnitude of the state can always grow, so that stabilization depends on the frequency of jump occurrence. From the assumptions on $\ell$, then $2\ell(r)r = r^2$ for $r\le 1$, and hence for $0 < a \le 1$, we have
  \begin{align*}
    \int_{\bar\alpha(a)}^a \frac{ds}{-\bar\varphi(s)} = \int_{\bar\alpha(a)}^a \frac{ds}{s^{2}}  = \frac{1}{\bar\alpha(a)} - \frac{1}{a} = \frac{1-2\delta(a)}{2\delta(a)a} \ge \frac{1}{2}.
  \end{align*}
  For $a\ge 1$, we have $\bar\alpha(a) = 1/2$, and hence
  \begin{align*}
    \int_{\bar\alpha(a)}^a \frac{ds}{-\bar\varphi(s)} \ge \int_{1/2}^a \frac{ds}{s^{2}} \ge \int_{1/2}^1 \frac{ds}{s^{2}} = 1.
  \end{align*}
  Therefore, if $\ell$ is such that, in addition, (\ref{eq:integralr}) is satisfied, then for all positive $\theta < 1/2$ it follows from Theorem~\ref{thm:sciss1}\ref{item:phind}) that the family of impulsive systems is strongly ISS if $\Gamma^* \subset \Gamma^\theta$. To see that (\ref{eq:integralr}) is unavoidable, suppose that $-\bar\varphi(s) = s^2$, so that everything remains as before excepting that (\ref{eq:integralr}) is not satisfied. Under zero input, $t_0 = 0$, $x(0) = 8$, the flow equation becomes $\dot x = x^2/2$ and its solution $x(t) = 2x(0)/[2 - x(0)t]$, with maximal (forward) interval of existence $[0,1/4)$. If $1/4 \le \theta < 1/2$, the solution may cease to exist before a jump occurs, and hence stability cannot be ensured, even if all the other assumptions of Theorem~\ref{thm:sciss1}\ref{item:phind}) are satisfied.
  \mer
\end{example}

\section{Sufficient Conditions for ISS of switched impulsive systems} \label{sec:switched}

In this section, we derive sufficient conditions for the $(h^o,h)$-ISS of switched impulsive systems with inputs. Let $\{\mathbf{f}_i\}_{i\in I_c}$ and $\{\mathbf{g}_j\}_{j\in I_d}$ be families of flow and jump maps, respectively, where $I_c$ and $I_d$ are index sets (i.e. arbitrary and nonempty). We assume that each flow map $\mathbf{f}_i$ satisfies the blanket assumption we make on the flow map $f$ in (\ref{eq:is}), in order to ensure (local, possibly nonunique) existence of solutions of the differential equation $\dot x(t)=\mathbf{f}_i(t,x(t),u(t))$ for each input $u\in \U$. Each triple $\sigma=(\{\tau^{\sigma}_k\}_{k=1}^N,\{i_k\}_{k=0}^N,\{j_k\}_{k=0}^N\}$ ---which we call impulsive and switching sequence---, with $\{\tau^{\sigma}_k\}_{k=1}^N =: \gamma_{\sigma} \in \Gamma$, $\{i_k\}_{k=0}^N\subset I_c$ and $\{j_k\}_{k=0}^N \subset I_d$, gives rise to the impulsive system with inputs $\Sigma_{\sigma} = (\gamma_{\sigma}, f_{\sigma}, g_{\sigma})$, with $f_{\sigma}(t,\xi,\mu) := \mathbf{f}_{\sigma_1(t)}(t,\xi,\mu)$ and $g_{\sigma}(t,\xi,\mu) := \mathbf{g}_{\sigma_2(t)}(t,\xi,\mu)$, where $\sigma_1:[0,\infty)\to I_c$ and $\sigma_2:[0,\infty)\to I_d$ are the switching signals \footnote{i.e. piecewise constant and right-continuous functions with a finite number of discontinuities in each compact interval.} defined by, respectively, $\sigma_1(t)=i_k$ and $\sigma_2(t)=j_k$ for all $t\in [\tau^{\sigma}_k, \tau_{k+1}^{\sigma})$, $0\le k <N+1$\footnote{Recall that $\tau^\sigma_0=0$ and that $\tau^{\sigma}_{N+1}=\infty$ if $N$ is finite}. Note that 
at each impulsive-switching time $\tau_k^{\sigma}$ the flow identified by the ``flow mode'' $\sigma_1({\tau_k^{\sigma}}^-)=i_{k-1}$ ends, the system jumps according to the jump map identified by the ``jump mode'' $\sigma_2(\tau_k^{\sigma})=j_k$, and then flow continues according to flow mode $\sigma_1({\tau_k^{\sigma}})=i_k$. 

We will address the weak or strong $(h^o,h)$-ISS of the systems $\Sigma_{\sigma}$ when $\sigma$ lives in some class of impulsive and switching sequences $\SW$. In order to apply the theory developed for parametrized families of impulsive systems to the study of the $(h^o,h)$-ISS of the systems $\Sigma_{\sigma}$, the impulsive and switching sequence $\sigma$ and the set $\SW$ will be regarded as, respectively, a parameter and a parameter set. In other words, $\sigma$ and $\SW$ will play the role of $\lambda$ and $\Lambda$ in the preceding sections. 

Given the index sets $I_c$, $I_d$, and a impulsive and switching sequence $\sigma$, it may be the case that not every combination of flow and jump modes is possible. Hence, one may want to describe classes of switching signals that incorporate this type of constraints. Therefore, let $J \subset I_c \times I_c \times I_d$ be nonempty. We write $\sigma\in\SW(J)$, if $(\sigma_1(t^-), \sigma_1(t), \sigma_2(t)) \in J$ for every $t \in \gamma_\sigma$. Without loss of generality, we consider that $I_c$, $I_d$ and $J$ satisfy $I_c = \{ i : (i,\hat\iota,j) \in J \text{ or }(\hat\iota,i,j)\in J \text{ for some }\hat\iota\in I_c,j\in I_d\}$ and $I_d = \{ j : (\hat\iota,i,j) \in J \text{ for some }\hat\iota,i\in I_c\}$. This just means that $I_c$ and $I_d$ do not contain modes that can never be used. We will require the following assumption.
\begin{as}\label{ass:issVsw}
 There exists a family of functions $\{\mathbf{v}_i\}_{i\in I_c}$, with $\mathbf{v}_i:\R_{\ge 0}\times \R^n\to \R$ locally Lipschitz for all $i\in I_c$, such that
 \begin{enumerate}[a)]
  \item there exist $\phi_1,\phi_2\in \Ki$ so that for all $i \in I_c$, $t\ge 0$ and $\xi \in \R^n$,\label{item:Vphi12sw}
    \begin{align} 
      \label{eq:bound1sw}
      \phi_1(h(t,\xi))\le \mathbf{v}_i(t,\xi)\le \phi_2(h^o(t,\xi));
    \end{align}
  \item there exist $\chi, \pi \in \Ki$, Carath\'eodory functions $\vphi_{i}:\R_{\ge 0}\times \R_{\ge 0}\to \R$, $i\in I_c$,  functions $\valpha_{\hat\iota,i,j}:\R_{\ge 0}\times \R_{\ge 0} \to \R_{\ge 0}$, $(\hat\iota,i,j)\in J$, such that the following hold for all $(\hat\iota,i,j)\in J$, $t\ge 0$, $\xi \in \R^n$ and $\mu \in \R^m$: \label{item:Vimplysw}
   \begin{enumerate}[i)]   
   \item \label{item:lyapboundsw}
   $D^{+}_{\mathbf{f}_{i}}\mathbf{v}_i(t,\xi,\mu) \le -\vphi_{i}(t,\mathbf{v}_{i}(t,\xi))$ if $\mathbf{v}_i(t,\xi)\ge \chi(|\mu|)$;
   \item \label{item:lyapbound2sw}
      $\mathbf{v}_{i}(t,\xi+\mathbf{g}_{j}(t,\xi,\mu)) \le \valpha_{\hat\iota,i,j}(t, \mathbf{v}_{\hat\iota}(t,\xi) )$ if $\mathbf{v}_{\hat\iota}(t,\xi)\ge \chi(|\mu|)$;
      \item \label{item:lyapbound3sw}
    $\mathbf{v}_{i}(t,\xi+\mathbf{g}_{j}(t,\xi,\mu))\le \pi(|\mu|)$ if  $\mathbf{v}_{\hat\iota}(t,\xi) \le \chi(|\mu|).$
  \end{enumerate}
  \end{enumerate}
\end{as}
In what follows, for each impulsive and switching sequence $\sigma$, we define the functions $V_{\sigma}(t,\xi) = \mathbf{v}_{\sigma_1(t)}(t,\xi)$, $\varphi_{\sigma}(t,r)=\vphi_{\sigma_1(t)}(t,r)$ and $\alpha_{\sigma}(t,r)=\valpha_{\hat\iota_k,i_k,j_k}(\tau_{k}^\sigma, r)$ if $t\in [\tau_k^{\sigma},\tau_{k+1}^{\sigma})$ and $k\ge 1$, where $\hat\iota_k=\sigma_1({\tau_k^{\sigma}}^-)$, $i_k=\sigma_1({\tau_k^{\sigma}})$ and $j_k=\sigma_2({\tau_k^{\sigma}})$.
Since the value of $\alpha_{\sigma}(t,r)$ for $t\in [0,\tau_1^k)$ is irrelevant, it could be arbitrarily defined; we define it as $\alpha_{\sigma}(t,r)=\valpha_{\hat\iota_1,i_1,j_1}(\tau_{1}^\sigma, r)$.
\begin{rem}\label{rem:vs}
Given a family of functions $\{\mathbf{v}_i\}_{i\in I_c}$ in the conditions of Assumption \ref{ass:issVsw} and an impulsive and switching signal $\sigma$, the family of impulsive systems $\{ \Sigma_{\sigma}\}_{\sigma \in \SW}$, with $\SW$ any family of impulsive and switching sequences,  together with the family of functions $\{V_{\sigma}\}_{\sigma \in \SW}$ satisfy Assumption \ref{ass:issV} with the same functions $\phi_1$, $\phi_2$, $\chi$ and $\pi$ appearing in Assumption~\ref{ass:issVsw}, and with $\varphi_\sigma$ and $\alpha_\sigma$ as defined above. \mer
\end{rem}
\begin{as} 
  \label{ass:issVexp} 
  There exist a partition\footnote{We allow that some member of a partition can be the empty set.} $I_c=I_c^s \cup I_c^n\cup I_c^u$, a partition $J = J^s \cup J^n \cup J^u$, and constants $c_s,c_u > 0$, $0 \le d_s < 1 < d_u$ such that Assumption~\ref{ass:issVsw} holds with $\vphi_i(t,r)=c_s\:r$ if $i\in I_c^s$, $\vphi_i(t,r)=0$ if $i\in I_c^n$, $\vphi_i(t,r)=-c_u\:r$ if $i\in I_c^u$, $\valpha_{\hat\iota,i,j}(t,r)=d_s\:r$ if $(\hat\iota,i,j)\in J^s$, $\valpha_{\hat\iota,i,j}(t,r)=r$ if $(\hat\iota,i,j)\in J^n$ and $\valpha_{\hat\iota,i,j}(t,r)=d_u\:r$ if $(\hat\iota,i,j)\in J^u$.
\end{as}

Under Assumption~\ref{ass:issVexp}, the right-hand sides of the first inequalities in items~\ref{item:lyapboundsw})--\ref{item:lyapbound2sw}) of Assumption~\ref{ass:issVsw} become linear in the Lyapunov-type functions, leading to exponential-type bounds. The sets $I_c^s$, $I_c^n$ and $I_c^u$ contain, respectively, the stabilizing, neutral and destabilizing flow modes. Analogously, $J^s$, $J^n$ and $J^u$ contain the jump-stabilizing, the jump-neutral and the jump-destabilizing combinations of modes.

For partitions $I_c=I_c^s \cup I_c^n \cup I_c^u$ and $J=J^s\cup J^n \cup J^u$ as in Assumption~\ref{ass:issVexp} and an impulsive and switching sequence $\sigma$, we define
\begin{align*}
  T^{\sigma,s}_{(t_0,t]} &:= \big|\{r \in (t_0,t]:\sigma_1(r)\in I_c^s\}\big|,\\
    T^{\sigma,n}_{(t_0,t]} &:= \big|\{r \in (t_0,t]:\sigma_1(r)\in I_c^n\}\big|,\\
  T^{\sigma,u}_{(t_0,t]} &:= \big|\{r \in (t_0,t]:\sigma_1(r)\in I_c^u\}\big|,\\
  \I_{(t_0,t]} &:= \left\{(\sigma_1(r^-),\sigma_1(r),\sigma_2(r)):r\in \gamma_{\sigma} \cap (t_0,t]\right\}, \\
  n^{\sigma,s}_{(t_0,t]} &:= \# \left[ \I_{(t_0,t]} \cap J^s \right],\quad   n^{\sigma,n}_{(t_0,t]} := \# \left[ \I_{(t_0,t]} \cap J^n \right],\\
  n^{\sigma,u}_{(t_0,t]}  &:= \# \left[ \I_{(t_0,t]} \cap J^u \right].
\end{align*}
Note that $T^{\sigma,s}_{(t_0,t]}+T^{\sigma,n}_{(t_0,t]} + T^{\sigma,u}_{(t_0,t]} = t-t_0$ and $n^{\sigma,s}_{(t_0,t]}+n^{\sigma,n}_{(t_0,t]} + n^{\sigma,u}_{(t_0,t]} = n^{\sigma}_{(t_0,t]} := n^{\gamma_{\sigma}}_{(t_0,t]}$. The quantities $T^{\sigma,s}_{(t_0,t]}$, $T^{\sigma,n}_{(t_0,t]}$ and $T^{\sigma,u}_{(t_0,t]}$ are, respectively, the total activation time of the modes in $I_c^s$, $I_c^n$ and $I_c^u$ in the interval $(t_0,t]$, while $n^{\sigma,s}_{(t_0,t]}$, $n^{\sigma,n}_{(t_0,t]}$ and $n^{\sigma,u}_{(t_0,t]}$ are the number of switching times $w$ in $(t_0,t]$ for which, respectively, $\alpha_{\sigma}(w,r)=d_s\:r$ , $\alpha_{\sigma}(w,r)=r$ and $\alpha_{\sigma}(w,r)=d_u\:r$.

\begin{teo} \label{thm:avdt-radtsw}
Let $\{\mathbf{f}_i\}_{i\in I_c}$ and $\{\mathbf{g}_j\}_{j\in I_d}$ be families of flow and jump maps, respectively, let $h^o,h \in \H$, let $J \subset I_c \times I_c \times I_d$, and consider a family of impulsive and switching sequences $\SW \subset \SW(J)$. Let Assumption \ref{ass:issVexp} hold. Then $\{\Sigma_\sigma\}_{\sigma \in \SW}$ is
  \begin{enumerate}[a)]
  \item \label{item:weaksw} weakly $(h^0,h)$-ISS if there exist $\eta,\mu > 0$ such that the following condition holds for every $\sigma \in \SW$:
    \begin{multline}
      \label{eq:hespanhasw}
      n^{\sigma,s}_{(t_0,t]}\ln(d_s)+n^{\sigma,u}_{(t_0,t]}\ln(d_u) - c_s T^{\sigma,s}_{(t_0,t]}\\ + c_u T^{\sigma,u}_{(t_0,t]}
       \le \mu-\eta(t-t_0), \quad \forall t\ge t_0 \ge 0;
    \end{multline}
  \item \label{item:sISS-adtsw}strongly $(h^o,h)$-ISS if there exist $\eta,\mu > 0$ such that the following condition holds for every $\sigma \in \SW$:
    \begin{multline}
      \label{eq:shespanhasw}
      n^{\sigma,s}_{(t_0,t]}\ln(d_s)+n^{\sigma,u}_{(t_0,t]}\ln(d_u) - c_s T^{\sigma,s}_{(t_0,t]} + c_u T^{\sigma,u}_{(t_0,t]}\\
       \le \mu-\eta\left[t-t_0+n^{\sigma}_{(t_0,t]}\right], \quad \forall t\ge t_0 \ge 0;      
      \end{multline}
  \end{enumerate}
\end{teo}
\begin{IEEEproof}
By virtue of Remarks~\ref{rem:equal} and~\ref{rem:vs}, Theorem~\ref{thm:main} and the fact that the functions $\alpha_{\sigma}(t,r)$ take the values $d_s r$, $r$ or $d_u r$, and are hence nondecreasing in $r$, we only have to prove that the family of comparison systems (\ref{eq:isce}) is weakly GUAS in case \ref{item:wISS-adt}), and strongly GUAS in case \ref{item:sISS-adtsw}). Let $\sigma \in \SW$, $t_0\ge 0$ and $w_0\in \R_{\ge 0}$. Taking into account that $\varphi_{\sigma}(t,r)=\phi_{\sigma}(t)\:r$, with $\phi_{\sigma}(t)=c_s$ if $\sigma_1(t)\in I_c^s$, $\phi_{\sigma}(t)=0$ if $\sigma_1(t)\in I_c^n$ and $\phi_{\sigma}(t) = -c_u$ if $\sigma_1(t)\in I_c^u$, and that for $\tau \in \gamma_{\sigma}$, $\alpha_{\sigma}(\tau,r)=d_s\:r$ if $(\sigma_1(\tau^-),\sigma_1(\tau),\sigma_2(\tau))\in J^s$, $\alpha_{\sigma}(\tau,r)=r$ if $(\sigma_1(\tau^-),\sigma_1(\tau),\sigma_2(\tau))\in J^n$  and $\alpha_{\sigma}(\tau,r)=d_u\:r$ if $(\sigma_1(\tau^-),\sigma_1(\tau),\sigma_2(\tau))\in J^u$, we have that the solution $w$ of  (\ref{eq:isce}) such that $w(t_0)=w_0$ satisfies for all $t\ge t_0$,
\begin{align*}
  w(t) &= w_0 e^{-\int_{t_0}^t \phi_{\sigma}(s)ds + n^{\sigma,s}_{(t_0,t]} \ln(d_s) + n^{\sigma,u}_{(t_0,t]} \ln(d_u)}.
\end{align*}
Since $\int_{t_0}^t \phi_{\sigma}(s)ds = c_s T^{\sigma,s}_{(t_0,t]} - c_u T^{\sigma,u}_{(t_0,t]}$, and using (\ref{eq:hespanhasw}) we obtain 
\begin{align*}
w(t)\le w_0 e^{\mu-\eta(t-t_0)}\quad \forall t\ge t_0,
\end{align*}
which shows that the family of comparison systems is weakly GUAS.
If (\ref{eq:shespanhasw}) is valid, it follows that
\begin{align*}
  w(t) &\le w_0 e^{\mu-\eta \left [t-t_0+n^{\sigma}_{(t_0,t]} \right ]} \quad \forall t\ge t_0,
\end{align*}
which shows that the family of comparison systems is strongly GUAS.
\end{IEEEproof}

By considering zero jump maps, i.e. $\mathbf{g}_j \equiv 0$ in Assumption~\ref{ass:issVexp}, Theorem~\ref{thm:avdt-radtsw} allows to recover or generalize well-known multiple-Lyapunov-function results for nonimpulsive switched systems. For instance, part (i) of Theorem 3.1 in \cite{vucha_auto07} can be straightforwardly derived from Theorem \ref{thm:avdt-radtsw}.
By contrast to other recent results for switched impulsive systems (e.g. \cite{liuliu_auto11,liuliu_scl12,lili_scl18}), Theorem~\ref{thm:avdt-radtsw} deals with flow modes and jump-combinations of modes that can be both stabilizing and destabilizing. Moreover, Theorems 1 to 3 in \cite{lili_scl18} become particular cases of Theorem~\ref{thm:avdt-radtsw} when restricted to the ISS case (i.e. $y=0$ in (1) of \cite{lili_scl18}) since the classes of impulsive switching sequences considered therein satisfy (\ref{eq:hespanhasw}).

\begin{example}
  Consider a second-order switched impulsive system with a single input having $I_c = I_d = \{1,2,3\}$, i.e. with 3 flow modes and 3 jump modes. Let the flow and jump maps be defined as
  \begin{align*}
    \mathbf{f}_1(t,\xi,\mu) &=
    \begin{bmatrix}
      -\xi_2\mu\\ \xi_1\mu
    \end{bmatrix}, &
    \mathbf{g}_1(t,\xi,\mu) &= 0\\ 
    \mathbf{f}_2(t,\xi,\mu) &=
    \begin{bmatrix}
      -\xi_1 + \mu\\ \xi_2 + |\xi|^2 \mu
    \end{bmatrix}, &
    \mathbf{g}_2(t,\xi,\mu) &=
                              \begin{bmatrix}
                                -1 & 0\\ 1 & -1
                              \end{bmatrix}\xi \\                          
    \mathbf{f}_3(t,\xi,\mu) &=
    \begin{bmatrix}
      \xi_1 + \mu\\ -\xi_1 + \xi_2 + \mu
    \end{bmatrix}, &
    \mathbf{g}_3(t,\xi,\mu) &=
                              \begin{bmatrix}
                                0 & 1\\ 0 & -1
                              \end{bmatrix}\xi  
  \end{align*}
  Let $J = \big\{(1,1,1)$, $(1,2,1)$, $(1,3,1)$, $(1,1,3)$, $(1,2,3)$, $(1,3,3)$, $(2,1,2)$, $(2,2,1)$, $(2,2,2)$, $(2,3,1)$, $(3,1,2)$, $(3,2,1)$, $(3,2,2)$, $(3,3,1)\big\}$, $h^o(t,\xi) = |\xi|$ and $h(t,\xi) = |\xi_1|$. We would like to assess the $(h^o,h)$-ISS of this system. We thus take $\mathbf{v}_1(t,\xi) = \frac{1}{2} |\xi|^2$, $\mathbf{v}_2(t,\xi) = \frac{1}{2} |\xi_1|^2$ and $\mathbf{v}_3 = \mathbf{v}_2$, $\chi(s) = 2s^2$ and $\pi = 2\chi$. These functions $\mathbf{v}_i$ satisfy (\ref{eq:bound1sw}) with $\phi_1(s) = \phi_2(s) = s^2/2$. By analyzing the flow and jump equations in relation to the requirements of Assumption~\ref{ass:issVexp}, the sets $I_c$ and $J$ may be partitioned as follows: $I_c^s = \{2\}$, $I_c^n = \{1\}$, $I_c^u = \{3\}$, with $c_s = 1$ and $c_u = 3$, $J^s = \{(2,2,2)$, $(3,2,2)\}$, $J^u = \{(1,1,3)$, $(1,2,3)$, $(1,3,3)\}$, and $J^n = J\setminus(J^s \cup J^u)$, with $d_s = 0$ and $d_u = 2$. Note that, e.g., the combination $(3,1,1)$ can never be included in the set $J$ if Assumption~\ref{ass:issVexp} is to be satisfied for the given functions $\mathbf{v}_i$. This means that we cannot allow to change from flow mode 3 to flow mode 1 using jump mode 1. This happens because $\mathbf{v}_3(t,\xi)$ is insensitive to the quantity $\xi_2$, the jump given by $\mathbf{g}_1$ leaves $\xi$ unchanged, and $\mathbf{v}_1(t,\xi)$ could be arbitrarily large even if $\mathbf{v}_3(t,\xi)$ is small. For illustration, we next derive sufficient conditions on $\SW \subset \SW(J)$ so that $\{\Sigma_\sigma\}_{\sigma\in\SW}$ is weakly or strongly $(h^o,h)$-ISS.

  a) $\SW$ is such that $\{ \gamma_\sigma : \sigma\in\SW \}$ is UIB and there exists $T>0$ such that $n^{\sigma,s}_{(t_0,t_0+T]} \ge 1$ for all $t_0 \ge 0$. For $0 \le t_0 < t < t_0 + T$, we have $n^{\sigma,s}_{(t_0,t]} \ln(d_s) + n^{\sigma,u}_{(t_0,t]} \ln(d_u) - c_s T^{\sigma,s}_{(t_0,t]} + c_u T^{\sigma,u}_{(t_0,t]} \le n^\sigma_{(t_0,t]} \ln(d_u) + c_u T \le \phi(T) \ln(2) + c_u T =: K$, where $\phi$ is the function given by the UIB property; hence, (\ref{eq:hespanhasw}) is satisfied with, e.g. $\mu = 2K$ and $\eta = K/T$. For $t\ge t_0 + T$, we have $n^{\sigma,s}_{(t_0,t]} \ge 1$ and since $d_s = 0$, then (\ref{eq:hespanhasw}) also is satisfied. By Theorem~\ref{thm:avdt-radtsw}, then weak $(h^o,h)$-ISS is ensured and by Proposition~\ref{prop:equivalence}, also strong $(h^o,h)$-ISS.

  b) $\SW$ is such that $n^{\sigma,s}_{(t_0,t]} \equiv 0$, i.e. stabilizing jumps do not occur. Suppose that there exist positive numbers $\tau_D$ and $T_0$, $N_0\in \N$ and $0<p_s\le 1$ so that $p_s>\frac{c_u\tau_D+\ln(d_u)}{(c_u+c_s)\tau_D}$ and that for every $\sigma \in \SW$ we have that for all $t\ge t_0\ge 0$, the number of destabilizing jumps and the total activation time of the stabilizing flow satisfy, respectively, $n^{\sigma,u}_{(t_0,t]}\le N_0+\frac{t-t_0}{\tau_D}$ and $T^{\sigma,s}_{(t,t_0]}\ge p_s(t-t_0)-T_0$. Then, for every $\sigma \in \SW$, (\ref{eq:hespanhasw}) holds with $\mu=N_0\ln(d_u)+c_sT_0$ and $\eta=p_s-\frac{c_u\tau_D+\ln(d_u)}{(c_u+c_s)\tau_D}>0$, and therefore the weak $(h^o,h)$-ISS of $\{\Sigma_{\sigma}\}_{\SW}$ is ensured by Theorem~\ref{thm:avdt-radtsw}. If, in addition, $\{ \gamma_\sigma : \sigma\in\SW \}$ is UIB, then $\{\Sigma_{\sigma}\}_{\SW}$ is strongly $(h^o,h)$-ISS of $\{\Sigma_{\sigma}\}_{\SW}$ by Proposition \ref{prop:equivalence}.
  \mer
\end{example}

Given $i\in I_c$ and $\theta>0$, we define $\S_{\textsc{dt}}^i[\theta]$ and $\S_{\textsc{rdt}}^i[\theta]$ as the sets of impulsive and switching sequences where each occurrence of flow mode $i$ has a dwell time of at least $\theta$ or at most $\theta$, respectively; i.e. $\sigma \in \S_{\textsc{dt}}^i[\theta]$ (resp. $\sigma \in \S_{\textsc{rdt}}^i[\theta]$) if $\tau^{\sigma}_{k+1}-\tau^{\sigma}_k\ge \theta$ ($\tau^{\sigma}_{k+1}-\tau^{\sigma}_k\le \theta$) for all $k$ for which $\sigma_1(\tau_k^{\sigma})=i$. We also define $\S_{\textsc{dt}}[\theta] = \cap_{i\in I_c} \S_{\textsc{dt}}^i[\theta]$ and  $\S_{\textsc{rdt}}[\theta]=\cap_{i\in I_c}\S_{\textsc{rdt}}^i[\theta]$, the sets of impulsive and switching sequences with dwell time at least $\theta$ and at most $\theta$, respectively. Note that an impulsive and switching sequence $\sigma$ satisfies $\sigma \in \S_{\textsc{dt}}[\theta]$ ($\sigma \in \S_{\textsc{rdt}}[\theta]$) if the sequence $\gamma_\sigma$ satisfies $\gamma_\sigma \in \Gamma_\theta$ ($\gamma_\sigma \in \Gamma^\theta$), with $\Gamma_\theta$ and $\Gamma^\theta$ as defined in Section~\ref{sec:suff-cond-iss}.

We also suppose the following.
\begin{as} 
  \label{ass:issVsw2}
  $I_c$ is finite and Assumption~\ref{ass:issVsw} holds with $\vphi_i(t,r)=p_i(t)\bar\vphi_i(r)$, $p_i$ locally integrable and nonnegative and $\bar\vphi_i$ continuous for all $i\in I_c$, and with $\valpha_{\hat\iota,i,j}(t,r) = \bar\valpha(r)$ for all $\hat\iota, i \in I_c$ and $j\in I_d$, $\bar\valpha$ continuous and positive definite.
\end{as}
\begin{teo} \label{thm:sw}
Let $\{\mathbf{f}_i\}_{i\in I_c}$ and $\{\mathbf{g}_j\}_{j\in I_d}$ be families of flow and jump maps, respectively, let $h^o, h \in \H$, and let $\SW$ be a family of impulsive and switching sequences. Let Assumption~\ref{ass:issVsw2} hold. Then $\{\Sigma_{\sigma}\}_{\sigma \in \SW}$ is strongly $(h^o,h)$-ISS if one of the following two conditions holds:
  \begin{enumerate}[a)]
  \item \label{item:phipdsw} for each $i\in I_c$, $\bar\vphi_i$ is positive definite and there exists a constant $\theta_i>0$ such that $\SW\subset \bigcap_{i\in I_c}\S_{\textsc{dt}}^i[\theta_i]$ and
    \begin{align}\label{eq:integral}
   \sup_{a>0} \int_a^{\bar\valpha(a)}\frac{ds}{\bar\vphi_i(s)}< \inf_{t\ge 0}\int_{t}^{t+\theta_i}p_i(s)\:ds=:M_i
    \end{align}
    with $M_i>0$;
  \item \label{item:phindsw}  for each $i\in I_c$, $\bar\vphi_i$ is negative definite and there exists a constant $\theta_i>0$ such that $\SW\subset \bigcap_{i\in I_c}\S_{\textsc{rdt}}^i[\theta_i]$ and
    \begin{align}
      \int_1^\infty \frac{ds}{-\bar\vphi_i(s)} &= \infty, \label{eq:integralrsw}\\ N_i^*:=\inf_{a>0} \int_{\bar\valpha(a)}^a\frac{ds}{-\bar\vphi_i(s)} &> \sup_{t\ge 0}\int_{t}^{t+\theta_i}p_i(s)\:ds=:M_i^*. \label{eq:integralr2sw}
    \end{align}
  \end{enumerate}
\end{teo}
The proof of Theorem~\ref{thm:sw} is given in Section~\ref{sec:pf-lem-comp}. Theorem~\ref{thm:sw} improves Theorems~4.1 and~4.2 of \cite{liuliu_scl12} by strengthening the conclusions on the one hand and by relaxing the assumptions 
on the other. In fact, in \cite[Theorems~4.1 and~4.2]{liuliu_scl12} only weak ISS results are given and the functions  $c_i$ and $g$, which play the roles of our $\bar\vphi_i$ and $\bar\valpha$ in part \ref{item:phipdsw}) and of $-\bar\vphi_i$ and $\bar\valpha$ in part \ref{item:phindsw}) are assumed of class $\Ki$ instead of merely positive definite. Relaxing the requirements on the functions involved leads to less conservative results. 


\section{Remaining proofs}
\label{sec:technical-proofs}

\subsection{Proof of Proposition \ref{prop:equivalence}}
\label{sec:proof-prop-equiv}

The proof requires the following result on $\KL$ functions.
\begin{lema}
  \label{lem:betapn}
  Let $\beta\in\KL$ and $\phi : \R_{\ge 0} \to \R_{\ge 0}$ be continuous and nondecreasing. Then, there exists $\hat\beta\in\KL$ such that
  \begin{align}
    \label{eq:betahat}
    \beta(r,s) &\le \hat\beta(r,s+\phi(s)),\quad \forall (r,s)\in\R_{\ge 0}^2.
  \end{align}
\end{lema}
\begin{IEEEproof}
  Since $\phi : \R_{\ge 0} \to \R_{\ge 0}$ is continuous and nondecreasing, there exists $\psi \in \ki$ such that $\phi(s) \le n_0+\psi(s)$ for all $s\ge 0$, where $n_0=\phi(0)$. Then for all $s\ge 0$, we have $s\ge \psi^{-1}\left (\max\{\phi(s)-n_0,0\}\right )$. Let $\varrho(s):=\psi^{-1}\left (\max\{s-n_0,0\}\right )$ for all $s\ge 0$. If $n_0\neq 0$,  pick any positive number $a$ and define $\hat\sigma \in \ki$ via $\hat \sigma(s)=\frac{a}{n_0} s$ if $0\le s <n_0$ and $\hat \sigma(s)=a+\varrho(s)$ otherwise. If $n_0=0$, consider $a=0$ and take $\hat \sigma=\varrho \in \ki$. Then it is easy to check that $\varrho(s)\ge \hat \sigma(s)-a$ for all $s\ge 0$. Pick any function $\sigma \in \Ki$ such that $\sigma(s)\le \hat \sigma(s)$ for all $s\ge 0$, $\sigma$ is differentiable on $(0,\infty)$ with derivative $\sigma'(s)\le 1$ for all $s>0$. For all $s> 0$, we have  
\begin{align}
  \label{eq:t-bound}
  s &\ge \varrho\left( \phi(s) \right) \ge 
          \sigma\left ( \phi(s) \right )-a,\\
  \label{eq:ine}
  s &\ge \sigma \left( s+\phi(s) \right) - \sigma \left( \phi(s) \right),
\end{align}
where (\ref{eq:ine}) follows from the fact that $\sigma'(s) \le 1$ for all $s>0$ implies that $\sigma(b) - \sigma(c) \le b - c$ for all $b>c\ge 0$.

Next, consider the given $\beta \in \KL$.
Due to Sontag's Lemma on class-$\KL$ functions (see, e.g. \cite{sontag_scl98}), there exist $\alpha_1,\alpha_2 \in \ki$ such that $\beta(r,s)\le \alpha_1(\alpha_2(r) e^{-s})$. Taking into account (\ref{eq:t-bound}) and (\ref{eq:ine}) it follows that 
\begin{align*}
\beta(r, s) &\le \beta \left (r,\frac{s}{2}+\frac{\sigma\left( \phi(s) \right) - a}{2} \right)\\
 &\le \alpha_1 \left ( \alpha_2(r)e^{a/2} \exp\left[-\dfrac{s+\sigma\left( \phi(s) \right) }{2}\right] \right)\\
& \le \alpha_1 \left ( \alpha_2(r) e^{a/2} e^{-\frac{\sigma \left( s+ \phi(s) \right)}{2}} \right).
\end{align*}
Defining $\hat\beta(r,s) =\alpha_1\left( \alpha_2(r) e^{\frac{a}{2}} e^{-\frac{\sigma(s)}{2}} \right)$, then $\hat\beta \in \KL$ and (\ref{eq:betahat}) follows.
\end{IEEEproof}	
\emph{Proof of Proposition~\ref{prop:equivalence}:}\\
($\Rightarrow$) That strong $(h^o,h)$-ISS implies weak ISS is straightforward, as explained at the beginning of Section~\ref{sec:relat-betw-stab}, so we need only prove the converse implication.\\
($\Leftarrow$) Let $\beta\in\KL$ characterize weak $(h^o,h)$-ISS, so that (\ref{eq:cwiss}) is satisfied, and let $\phi : \R_{\ge 0} \to \R_{\ge 0}$ be the continuous and nondecreasing function as per the UIB property. Let $\hat\beta\in\KL$ be given by Lemma~\ref{lem:betapn}. Then,
\begin{align}
  \label{eq:betauib}
  \beta(r,s) &\le \hat\beta(r,s+\phi(s)) \le \hat\beta(r,s+n^\gamma_{(t_0,t_0+s]}),
\end{align}
for all $r,s,t_0\ge 0$ and all $\gamma\in\S$, where the last inequality above follows from $n^\gamma_{(t_0,t_0+s]} \le \phi(s)$ and $\hat\beta\in\KL$. Replacing $r = h^o(t_0,x(t_0))$, $s=t-t_0$, and recalling (\ref{eq:cwiss}), then (\ref{eq:ciss}) follows and strong $(h^o,h)$-ISS is established.$\hfill$
\QED

\subsection{Proof of Theorem~\ref{thm:sw}}
\label{sec:pf-lem-comp}
The proof of the theorem requires the following result.
\begin{lema} \label{lem:kinf} Let $\varphi:\R_{\ge 0}\to \R_{\ge 0}$ and $\alpha:\R_{\ge 0}\to \R_{\ge 0}$ be continuous and positive definite, let $\phi:\R_{\ge 0}\to \R_{\ge 0}$ be locally integrable and let $\theta>0$. Consider the following conditions:
 \begin{enumerate}[a)]
  \item \label{item:akinf} 
  \begin{align} \sup_{a>0}\int_a^{\alpha(a)}\frac{ds}{\varphi(s)}<\inf_{t\ge 0} \int_{t}^{t+\theta} \phi(s)\:ds =: M, \label{eq:ca}
   \end{align}
with $M>0$;
\item \label{item:bkinf}
 \begin{align} \int_{1}^{\infty}\frac{ds}{\varphi(s)} & =\infty \label{eq:1}\\
 \inf_{a>0}\int_{\alpha(a)}^a\frac{ds}{\varphi(s)}& >\sup_{t\ge 0} \int_{t}^{t+\theta} \phi(s)\:ds. \label{eq:cb}
   \end{align}
 \end{enumerate}
 Then, there exists $\bar \alpha\in \Ki$ with $\alpha\le \bar \alpha$, such that (\ref{eq:ca}) or (\ref{eq:cb}) hold, with $\bar \alpha$ in place of $\alpha$, if, respectively, \ref{item:akinf}) or \ref{item:bkinf}) hold.  
\end{lema}
\begin{IEEEproof} 
 Suppose \ref{item:akinf}) holds. Let $N=\sup_{a>0}\int_a^{\alpha(a)}\frac{ds}{\varphi(s)}$. Since $N<M$ and $M>0$, we can pick a number $M^*>0$ such that $N<M^*<M$. For $a>0$ define \begin{align*}
          \hat \alpha(a)=\sup\left \{r\ge a:\: \int_a^{r}\frac{ds}{\varphi(s)}\le M^*\right \}.                                                                                                                                                                                                                                                                                                                                                \end{align*}
Note that $\hat \alpha(a)=\infty$ if $\int_{a}^{\infty} \frac{ds}{\varphi(s)}\le M^*$ and that $\hat \alpha(a)$ is finite, $\hat \alpha(a)>a$ and $\int_a^{\hat \alpha(a)}\frac{ds}{\varphi(s)}= M^*$ when $\int_{a}^{\infty} \frac{ds}{\varphi(s)}> M^*$. Also note that $\alpha(a)<\hat \alpha(a)$ for all $a>0$, since $\int_a^{\alpha(a)}\frac{ds}{\varphi(s)}\le N<M^*=\int_a^{\hat \alpha(a)}\frac{ds}{\varphi(s)}$ when $\hat \alpha(a)$ is finite. Two cases are possible.

\emph{Case 1:} $\hat \alpha(a)=\infty$ for all $a>0$. In this case we take any $\bar \alpha \in \Ki$ such that $\alpha\le \bar \alpha$ and $\bar \alpha(a)>a$ for all $a>0$. Since $\int_a^{\bar \alpha(a)}\frac{ds}{\varphi(s)} \le \int_a^{\infty}\frac{ds}{\varphi(s)}\le M^*$ for all $a>0$, (\ref{eq:ca}) holds with $\bar \alpha$ in place of $\alpha$.

\emph{Case 2:} $\hat \alpha(a)$ is finite for some $a>0$. Let $b=\sup\{a>0:\:\hat \alpha(a)<\infty\}$. If $b=\infty$, then it follows that $\int_{a}^{\infty} \frac{ds}{\varphi(s)}> M^*$ for all $a>0$, which in turn implies that $\int_{a}^{\infty} \frac{ds}{\varphi(s)}=\infty$ for all $a>0$. Define $F(r)=\int_{1}^{r} \frac{ds}{\varphi(s)}$ for $r>0$. $F$ is strictly increasing, continuous and $\lim_{r\to \infty}F(r)=\infty$. Let $c=\lim_{r\to 0^+}F(r)$. Then there exists $F^{-1}:(c,\infty)\to (0,\infty)$ and it is continuous and strictly increasing. Since $F(\hat \alpha(a))-F(a)=\int_a^{\hat \alpha(a)}\frac{ds}{\varphi(s)}= M^*$, it follows that $\hat \alpha(a)=F^{-1}(F(a)+M^*)$. Hence $\hat \alpha$ is strictly increasing, continuous, $\hat \alpha(a)>a$ for all $a>0$ and $\lim_{a \to \infty}\hat \alpha(a)=\infty$. In addition, taking into account that $\alpha(a)<\hat \alpha(a)$ for all $a>0$, then there exists $\bar\alpha \in \Ki$ such that $\max\{\alpha(a),a\}<\bar \alpha(a)\le \hat \alpha(a)$ for all $a>0$. That (\ref{eq:ca}) holds with $\bar \alpha$ in place of $\alpha$ follows from the following inequalities
\begin{align*}
\int_a^{\bar \alpha(a)}\frac{ds}{\varphi(s)}
\le  \int_a^{\hat \alpha(a)}\frac{ds}{\varphi(s)} \le M^*.
\end{align*}

If $b$ is finite, then $\int_b^{\infty}\frac{ds}{\varphi(s)}=M^*$. In consequence $\hat \alpha(a)=\infty$ for all $a\ge b$ and $\hat \alpha(a)$ is finite and $\int_a^{\hat \alpha(a)}\frac{ds}{\varphi(s)}=M^*$ for all $0<a<b$. Defining $F:(0,\infty)\to \R$ and $c$ as above, and considering $d=\lim_{r\to \infty}F(r)=F(b)+M^*$, we have that there exists $F^{-1}:(c,d)\to (0,\infty)$ and that it is continuous and strictly increasing. In addition $\hat \alpha(a)=F^{-1}(F(a)+M^*)$ for all $0<a<b$. So, $\hat \alpha$ is continuous, strictly increasing, $\lim_{a\to b^-}\hat \alpha(a)=\infty$ and $\id <\hat \alpha$ on $(0,b)$, where $\id$ is the identity function. Taking also into account that $\alpha<\hat\alpha$ on $(0,\infty)$ it follows the existence of $\bar \alpha\in \Ki$ such that $\max\{\alpha(a),a\}<\bar \alpha(a)\le \hat \alpha(a)$ for all $a>0$. That (\ref{eq:ca}) holds with $\bar \alpha$ in lieu of $\alpha$ can be proved as in the case $b=\infty$.

Suppose now that \ref{item:bkinf}) holds. Let $N^*=\inf_{a>0} \int_{\alpha(a)}^a \frac{ds}{\varphi(s)}$. Note that $N^*>0$. Consider the continuous and strictly increasing function $F:(0,\infty)\to \R$ defined as above. Then (\ref{eq:1}) implies that $\lim_{r\to \infty}F(r)=\infty$. Suppose for a contradiction that $\lim_{r\to 0^+}F(r)=c$ with $c\in \R$. Then, 
\begin{align} 
N^*\le \lim_{a\to 0^+}\int_{\alpha(a)}^a \frac{ds}{\varphi(s)}=\lim_{a\to 0^+}F(a)-F(\alpha(a))=0, 
\end{align}
which is absurd. Therefore $\lim_{r\to 0^+}F(r)=-\infty$ and there exists $F^{-1}:(-\infty,\infty)\to (0,\infty)$. Define $\bar \alpha(a)=F^{-1}(F(a)-N^*)$ for $a>0$ and $\bar \alpha(0)=0$. It follows that $\bar \alpha \in \Ki$ and
\begin{align*}
\int_{\bar \alpha(a)}^a \frac{ds}{\varphi(s)}=N^*\le \int_{\alpha(a)}^a \frac{ds}{\varphi(s)}\quad \forall a>0.
\end{align*}
The latter implies that (\ref{eq:cb}) holds with $\bar \alpha$ in place of $\alpha$ and that $\alpha\le \bar \alpha$.
\end{IEEEproof}

\begin{IEEEproof}[Proof of Theorem~\ref{thm:sw}]
  Suppose that $\bar\vphi_i$, $p_i$ and $\bar\valpha$ and $\SW$ satisfy
  \ref{item:phipdsw}) of Theorem~\ref{thm:sw}. Due to Lemma \ref{lem:kinf}, without loss of generality we can suppose that the function $\bar\valpha$ belongs to $\Ki$. In fact, in case $\bar\valpha$ is only positive definite, for each $i\in I_c$, by applying Lemma \ref{lem:kinf} with $\varphi=\bar\vphi_i$, $\phi=p_i$ and $\theta=\theta_i$, there must exist $\bar\alpha_i\in \Ki$ such that $\bar \valpha\le \bar \alpha_i$ and (\ref{eq:integral}) hold with $\bar \alpha_i$ instead of $\bar \valpha$. Then, $\bar\alpha = \min_{i\in I_c}\bar \alpha_i \in \Ki$, $\bar\valpha \le \bar \alpha$ and (\ref{eq:integral}) holds with $\bar \alpha$ in place of $\bar \valpha$ for all $i\in I_c$. In consequence, the hypotheses of Theorem~\ref{thm:sw} remain valid if we replace $\bar\valpha$ by $\bar\alpha$. 
  
  By virtue of Remark~\ref{rem:vs} and Theorem~\ref{thm:main}, the weak $(h^0,h)$-ISS of
  $\{\Sigma_{\sigma}\}_{\sigma \SW}$ is established provided the
 family of comparison systems (\ref{eq:isc}), with
 $\varphi_{\sigma}(t,r)=p_{\sigma_1(t)}(t)\bar\vphi_{\sigma_1(t)}(r)$,
  $\alpha_{\sigma}(t,r)=\bar\valpha(r)$, $\sigma \in \SW$, is weakly GUAS. Since $\alpha_{\sigma}(t,r)$ is increasing in $r$ for all $\sigma \in \SW$, due to Remark \ref{rem:equal} the weak GUAS of the comparison systems (\ref{eq:isc}) follows from that of the systems (\ref{eq:isce}), which, in turn, can be obtained following the same steps of the proof of Lemma 3.2 in \cite{liuliu_scl12}. We note that the proof of that lemma remains valid if the functions $\alpha_i$ considered there are assumed continuous and positive definite instead of locally Lipschitz and of class $\ki$.

Once we have proven that $\{\Sigma_{\sigma}\}_{\sigma \in \SW}$ is weakly $(h^o,h)$-ISS, then strong $(h^o,h)$-ISS follows from Proposition~\ref{prop:equivalence} and the fact that the set of impulse times is UIB, since $\SW \subset \bigcap_{i\in I_c} \S_{\textsc{dt}}^i[\theta_i] \subset \S_{\textsc{dt}}[\underline{\theta}]$, where $\underline{\theta} := \min_{i\in I_c} \theta_i > 0$.

		
  Suppose now that $\bar\vphi_i$, $p_i$, $\bar\valpha$ and $\SW$ satisfy~\ref{item:phindsw}) of Theorem~\ref{thm:sw}. As in the previous case, by applying Lemma~\ref{lem:kinf} we can assume without loss of generality that $\bar\valpha\in\Ki$. By virtue of Remarks~\ref{rem:vs} and \ref{rem:equal}  and Theorem \ref{thm:main}, the strong $(h^0,h)$-ISS of $\{\Sigma_{\sigma}\}_{\sigma \SW}$ is established provided the family of comparison systems (\ref{eq:isce}), with $\varphi_{\sigma}(t,r) = p_{\sigma_1(t)}(t) \bar\vphi_{\sigma_1(t)}(r)$, $\alpha_{\sigma}(t,r) = \bar\valpha(r)$ and $\sigma \in \SW$ is strongly GUAS. We require the following claims.

  \emph{Claim 1:} Let $i\in I_c$. For each $t_0\ge 0$ and $\zeta\ge 0$
  there exists a unique forward-in-time solution
  $w_i(\cdot,t_0,\zeta):[t_0,\infty)\to \R_{\ge 0}$ of the initial value
  problem $\dot w(t) = -p_i(t) \bar\vphi_i(w(t))$, $w(t_0)=\zeta$. In addition, there exists $\nu_i\in \Ki$ so that $w_i(t,t_0,\zeta)\le \nu_i(\zeta)$ for
  all $t\in [t_0,t_0+\theta_i]$.

  \emph{Proof:} Define $F_i:(0,\infty)\to \R$ by
  \begin{align}
    \label{eq:defFi}
    F_i(r) &= \int_{1}^r \frac{ds}{-\bar\vphi_i(s)}.
  \end{align}
  From the last part of the proof of Lemma \ref{lem:kinf}, with $-\bar \vphi_i$ instead of $\varphi$, we have that the function $F_i$ has an inverse $F^{-1}_i:(-\infty,\infty)\to (0,\infty)$ which is continuous and strictly increasing. From the existence of $F_i^{-1}$ it is a simple exercise to show that the initial value problem $\dot w(t)=-p_i(t)\bar\vphi_i(w(t))$, $w(t_0)=\zeta$, with $\zeta\ge 0$ has a unique forward-in-time solution $w_i(\cdot,t_0,\zeta)$ defined for all $t\ge t_0$, which is given by the formula $w_i(t,t_0,\zeta) = F_i^{-1}\left( F_i(\zeta) + \int_{t_0}^t p_i(s)\:ds \right)$ if $\zeta>0$ and $w_i(t,t_0,\zeta)=0$ if $\zeta=0$. Define $\nu_i : \R_{\ge 0} \to \R_{\ge 0}$ via $\nu_i(r)=F_i^{-1}(F_i(r)+M_i^*)$ if $r > 0$ and $\nu_i(0)=0$. The function $\nu_i$ is continuous, increasing, $\lim_{r\to \infty}\nu_i(r)=\infty$ and $\lim_{r\to 0^+}\nu_i(r)=0=\nu_i(0)$. Then, $\nu_i \in \Ki$. We also have that $w_i(t,t_0,\zeta)\le \nu(\zeta)$ for all $t\in [t_0,t_0+\theta_i]$, because $\int_{t_0}^{t_0+\theta_i}p_i(s)\:ds\le M_i^*$.\mer

  \emph{Claim 2:} Let $i\in I_c$ and let $F_i$ be as defined in (\ref{eq:defFi}). Define $G_i(\zeta)=F_i^{-1}(F_i(\bar\valpha(\zeta))+M_i^*)$ for $\zeta>0$
  and $G_i(0)=0$. Then $G_i$ is continuous and $\zeta-G_i(\zeta)$
  is positive definite. In addition
  $w_i(t,t_0,\bar\valpha(\zeta))\le G_i(\zeta)$ for all
  $t\in [t_0,t_0+\theta_i]$.

  \emph{Proof:} The continuity of $G_i$ at $\zeta>0$ follows from the
  continuity of the functions involved in its definition. The facts
  that $F_i(\bar\valpha(\zeta))\to -\infty$ as $\zeta \to 0^+$ and
  that $F_i^{-1}(s)\to 0$ as $s\to -\infty$ imply that $G$ is
  also continuous at $0$. Let $\zeta>0$. From the definitions of $G_i$
  and $F_i$ we have that
  \begin{align*}
    \int_1^{G_i(\zeta)}\frac{ds}{-\bar\vphi_i(s)}=\int_1^{\bar\valpha(\zeta)} \frac{ds}{-\bar\vphi_i(s)}+M_i^*.
  \end{align*}
  Therefore,
  \begin{align*}
    M_i^*=\int_{\bar\valpha(\zeta)}^{G_i(\zeta)} \frac{ds}{-\bar\vphi_i(s)} < N_i^*\le \int_{\bar\valpha(\zeta)}^{\zeta} \frac{ds}{-\bar\vphi_i(s)}.
  \end{align*}
  In consequence,
  \begin{align*}
    F_i(\zeta)-F_i(G_i(\zeta))=\int_{G_i(\zeta)}^{\zeta} \frac{ds}{-\bar\vphi_i(s)} \ge N_i^*-M_i^*>0.
  \end{align*}
  Thus $\zeta-G_i(\zeta)>0$. Finally,
  \begin{align*}
    w_i(t,t_0,\bar\valpha(\zeta)) &= F_i^{-1}\left [F_i (\bar\valpha(\zeta)) + \int_{t_0}^t p_i(s)\:ds \right]\le G_i(\zeta)
  \end{align*}
  since $\int_{t_0}^t p_i(s)\:ds\le M_i^*$ for all $t\in [t_0,t_0+\theta_i]$.\mer

  Let $G=\max_{i\in I_c}G_i$ and $\nu =\max_{i\in I_c}\nu_i$. Then $G$ is continuous, $\zeta-G(\zeta)$ is positive definite and $\nu \in \Ki$. Consider the difference inclusion
  \begin{align}\label{eq:di}
    \zeta_{k+1}\in H(\zeta_k):= [0,G(|\zeta_k|)].
  \end{align}
  Since $W(\zeta)=|\zeta|$ is a Lyapunov function for (\ref{eq:di}),
  because $W(\xi)-W(\zeta)\le -[|\zeta|-G(|\zeta|)]$ for all
  $\xi \in H(\zeta)$ and $|\zeta|-G(|\zeta|)$ is a positive definite
  function, the difference inclusion (\ref{eq:di}) is GUAS (see
  \cite{keltee_scl04}). In consequence, there exists
  $\tilde\beta \in \KL$ so that for every solution
  $\{\zeta_k\}_{k=0}^\infty$ of (\ref{eq:di}), then
  \begin{align} \label{eq:betatilde} |\zeta_k|\le \tilde
    \beta(|\zeta_0|,k)\quad \forall k\in \N_0.
  \end{align}
  Let $\sigma \in \SW$ and let $w:[t_0,T_w)\to \R_{\ge 0}$ be a
  solution of (\ref{eq:isce}), with $\lambda=\sigma$,
  $\gamma_\lambda=\gamma_{\sigma}=\{\tau_k\}_{k=1}^N$, initial time
  $t_0\ge 0$ and initial condition $\zeta_0\ge 0$. Let $t\in [t_0,T_w)$
  and let $\ell=n^{\sigma}_{(t_0,t]}$. If $\ell=0$, then
  $t-t_0\le \theta_{i_0}$, where $i_0=\sigma_1(t_0)$. By using
   Claim 1, it follows that
  \begin{align} \label{eq:z0} w(t)= w_{i_0}(t,t_0,\zeta_0)\le
    \nu_{i_0}(\zeta_0)\le \nu(\zeta_0).
  \end{align}
  If $\ell=1$, then there is just one impulse time $\tau_{k_1}$ in
  $(t_0,t]$. Define $\zeta_1=w(\tau_{k_1}^-)$, $i_0=\sigma_1(t_0)$ and
  $i_1=\sigma_1(\tau_{k_1})$. Then, since
  $t-\tau_{k_1}\le \theta_{i_1}$ and $\tau_{k_1}-t_0\le \theta_{i_0}$
  and using Claims 1 and 2, we have
  \begin{align}
    w(t)&= w_{i_1}(t,\tau_{k_1},\bar\valpha(\zeta_1))\le G_{i_1}(\zeta_1)\le \zeta_1 \notag\\
    \label{eq:z1}
        &= w_{i_0}(\tau_{k_1},t_0,\zeta_0)\le \nu_{i_0}(\zeta_0)\le \nu(\zeta_0).
  \end{align}		
  If $\ell>1$ then there are exactly $\ell$ impulse times
  $t_0<\tau_{k_1}<\cdots<\tau_{k_{\ell}}\le t$, where
  $\tau_{k_j}=\tau_{{k_1}+j-1}$. Let $i_0=\sigma_1(t_0)$ and let
  $i_j=\sigma_1(\tau_{k_j})$ and $\zeta_j=w(\tau_{k_j}^-)$ for
  $j=1,\ldots,\ell$. Then
  \begin{align*}
    \zeta_1 &= w_{i_0}(\tau_{k_1},t_0,\zeta_0)\le \nu_{i_0}(\zeta_0),\\
    \zeta_2 &=  w_{i_1}(\tau_{k_2},\tau_{k_1},\bar\valpha(\zeta_1))\le  G_{i_1}(\zeta_1)\le G(\zeta_1),\\
    \zeta_3 &=  w_{i_2}(\tau_{k_3},\tau_{k_2},\bar\valpha(\zeta_2))\le  G_{i_2}(\zeta_2)\le G(\zeta_2),
  \end{align*}
  and, in general, $\zeta_{j+1}\le G(\zeta_j)$ for $j=1,\ldots,\ell-1$. By
  defining $\zeta_j=0$ for all $j\ge \ell+1$ it follows that
  $\{z_{k}=\zeta_{k+1}\}_{k=0}^{\infty}$ is a solution of
  (\ref{eq:di}). So $\zeta_{\ell}\le \tilde \beta(\zeta_1,\ell-1)$ and then
  \begin{align}
    w(t) &= w_{i_{\ell}}(t,\tau_{k_{\ell}},\bar\valpha(\zeta_{\ell}))
           \le G_{i_{\ell}}(\zeta_{\ell})\le \zeta_{\ell} \notag \\ \label{eq:zl} &\le \tilde \beta(\zeta_1,\ell-1)\le \tilde\beta(\nu(\zeta_0),\ell-1).
  \end{align}
  Let $\hat \beta \in \KL$ be defined as follows
  \begin{align}
    \hat\beta(r,t)=\begin{cases} (2-t)\tilde\beta(\nu(r),0)\quad r\ge 0, \; 0\le t\le 1, \\
      \tilde \beta(\nu(r),t-1)\quad \quad \; r\ge 0,\; t> 1. 
    \end{cases}
  \end{align}
  Then, from (\ref{eq:z0}), (\ref{eq:z1}), (\ref{eq:zl}) and the
  definition of $\hat\beta$, we have
  \begin{align} \label{eq:zt} w(t)\le \hat
    \beta\left(\zeta_0,n^\sigma_{(t_0,t]} \right)\quad \forall t\in [t_0,T_w).
  \end{align}
  Since $\SW \subset \cap_{i\in I_c} \S_{\textsc{rdt}}^i[\theta_i]$ then
  $n^\sigma_{(t_0,t]} \ge \frac{t-t_0}{\theta}-1 \ge
  \frac{t-t_0}{\theta_1}-1$,
  with $\theta=\max_{i\in I_c}\theta_i$ and
  $\theta_1 = \max\{\theta,1\}$. In consequence,
  \begin{align*}
    n^\sigma_{(t_0,t]} &\ge \frac{t-t_0}{2\theta} + \frac{n^\sigma_{(t_0,t]}}{2} - \frac{1}{2} \ge \frac{t-t_0+n^\sigma_{(t_0,t]}}{2\theta_1} - \frac{1}{2}.
  \end{align*}
  Let $\alpha_1,\alpha_2\in\Ki$ be given by Sontag's Lemma so that
  $\hat\beta(r,s) \le \alpha_1(\alpha_2(r) e^{-s})$ for all
  $r,s \ge 0$. Then,
  \begin{align}
    \hat\beta(\zeta_0,n^\sigma_{(t_0,t]}) &\le \hat\beta\left(\zeta_0, \max\left\{ \frac{t-t_0 + n^{\sigma}_{(t_0,t]}}{2\theta_1} - \frac{1}{2}, 0 \right\} \right)\notag\\
                                      &\le \alpha_1\left( \alpha_2(\zeta_0) e^{-\max\left\{ \frac{t-t_0 + n^\sigma_{(t_0,t]}}{2\theta_1} - \frac{1}{2}, 0 \right\}} \right)\notag\\
    \label{eq:hatbetaineq}
                                      &\le \alpha_1\left( \alpha_2(\zeta_0) e^{1/2}e^{-\frac{t-t_0 + n^\sigma_{(t_0,t]}}{2\theta_1}} \right).
  \end{align}
  Define $\beta\in\KL$ via
  \begin{align*}
    \beta(r,s) &:= \alpha_1\left( \alpha_2(r) e^{1/2} e^{\frac{-s}{2\theta_1}} \right).
  \end{align*}
  From~(\ref{eq:zt}) and (\ref{eq:hatbetaineq}), we finally obtain
  \begin{align} \label{eq:zt2} w(t) &\le
    \beta\left( \zeta_0, t-t_0 + n^\sigma_{(t_0,t]}\right)\quad \forall t\in
    [t_0,T_w),
  \end{align} 
  which shows that (\ref{eq:isce}), and hence (\ref{eq:isc}), is strongly GUAS. The result follows by application of Theorem~\ref{thm:main}.
\end{IEEEproof}

\subsection{Proof of Theorem~\ref{thm:sciss1}}
\label{sec:pf-thm-sciss1}

 By assumption, $\varphi_\lambda$ and $\alpha_\lambda$ are independent of $\lambda$, then (\ref{eq:isc}) consists in the family of comparison systems 
\begin{subequations}
  \label{eq:isc2}
  \begin{align}
    \label{eq:is-ctc2}
    \dot{z}(t) &\in(-\infty,-\phi(t)\bar \varphi(z(t))], & 
    t\notin \gamma_{\lambda},    \displaybreak[0] \\
    \label{eq:is-stc2}
    z(t) &\in[0,\bar \alpha(z(t^-))], & 
    t\in \gamma_{\lambda},
  \end{align}
\end{subequations}
with $\gamma_{\lambda} \in \Gamma_{\Lambda}$.

Such a family of comparison systems can be seen as those arising in Theorem~\ref{thm:sw} when its assumptions are satisfied with $\bar\vphi_i(r) = \bar\varphi(r)$, $p_i(t)=\phi(t)$ and $\theta_i=\theta$ for all $i\in I_c$ and $\bar\valpha(r) = \bar\alpha(r)$, and the set of impulsive and switching sequences $\SW$ is a subset of $\S_{\textsc{dt}}[\theta]$ when  \ref{item:phipd}) of Theorem \ref{thm:sciss1} holds and of $\S_{\textsc{rdt}}[\theta]$ when \ref{item:phind}) of such a theorem holds. So, following the steps of the first part of the proof of Theorem \ref{thm:sw} we have that (\ref{eq:isc2}) is weakly GUAS when \ref{item:phipd}) of Theorem \ref{thm:sciss1} holds, and therefore $\Sigma_{\Lambda}$ is weakly $(h^o,h)$-ISS due to Theorem \ref{thm:main}. That it is strongly $(h^o,h)$-ISS follows from Proposition \ref{prop:equivalence} and the fact that $\Gamma_{\Lambda}$ is UIB. In case \ref{item:phind}) of Theorem \ref{thm:sciss1} holds, following the steps in the second part of the proof of Theorem \ref{thm:sw} we can conclude that (\ref{eq:isc2}) is strongly GUAS and then that $\Sigma_{\Lambda}$ is strongly $(h^o,h)$-ISS due to Theorem \ref{thm:main}. \qed



\section{Conclusions}
\label{sec:conclusions}

We have provided a Lyapunov-type method for establishing uniform ISS of families of time-varying impulsive systems and shown how the given results generalize existing results for impulsive and nonimpulsive, switched and nonswitched systems. We have addressed weak and strong ISS: the decaying term in weak ISS is insensitive to the occurrence of jumps whereas that of strong ISS causes additional decay whenever a jump occurs. To allow greater generality, our results were given in the (time-varying) two-measure framework. Future work may be aimed at providing converse Lyapunov theorems and hence assessing the degree to which the given conditions are only sufficient for ISS or whether they may become also necessary.



\bibliographystyle{IEEEtran}
\bibliography{/home/hhaimo/latex/strings.bib,/home/hhaimo/latex/complete_v2.bib,/home/hhaimo/latex/Publications/hernan_v2.bib}

\begin{biography}[{\includegraphics[width=1in,height=!]{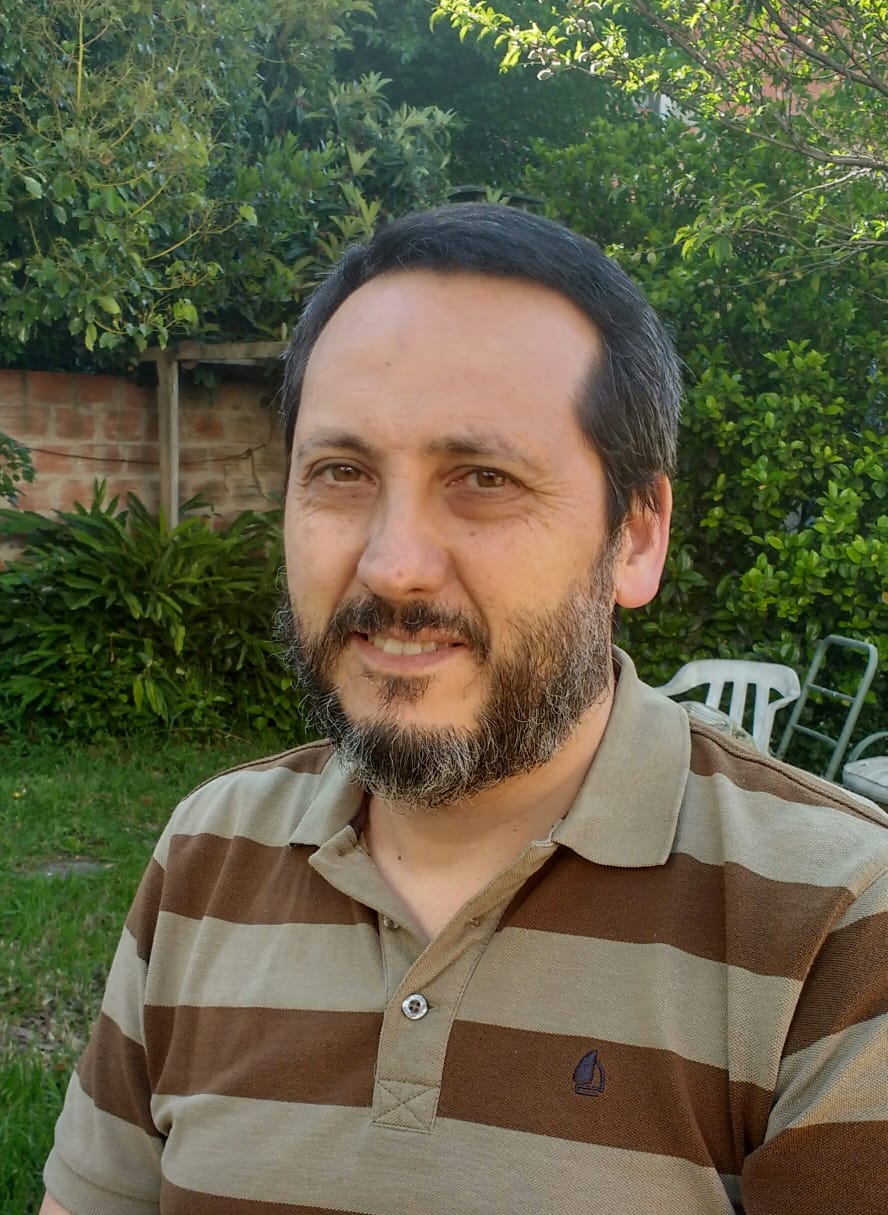}}]
	{Jos\'e Luis Mancilla Aguilar} received the Licenciado en Matem\'atica degree and the Doctor’s degree in mathematics from the Universidad Nacional de Buenos Aires (UBA), Argentina, in 1994 and 2001, respectively. From 1993 to 1995, he received a Research
Fellowship from the Argentine Atomic Energy Commission (CNEA) in nonlinear control. Since 1995, he has been with the Department of Mathematics of the Facultad de Ingenier\'{\i}a (UBA), where he is currently a part-time Associate Professor. Since 2005, he has held a Professor position at the Department of Mathematics of the Instituto Tecnol\'ogico de Buenos Aires (ITBA) and currently is the head of the Centro de Sistemas y Control (CeSyC). His research interests include hybrid systems and nonlinear control.
\end{biography}

\begin{biography}[{\includegraphics[width=1in,height=!]{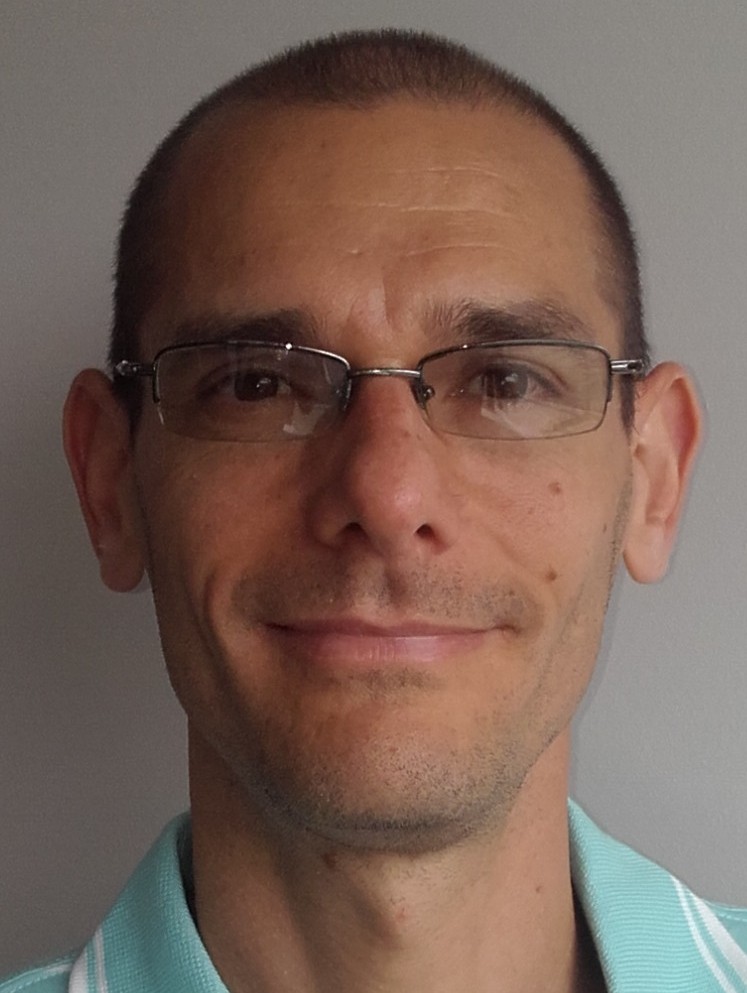}}]
	{Hernan Haimovich} received the Electronics Engineering degree with highest honours in 2001 from the Universidad Nacional de Rosario (UNR), Argentina, and the Ph.D. degree from The University of Newcastle, Australia, in 2006. In 2006, he worked as a Research Assistant at the Centre for Complex Dynamic Systems and Control at the University of Newcastle, Australia, and later as an Argentine Research Council (CONICET) Postdoctoral Research Fellow at the UNR, Argentina. Since 2007, Dr. Haimovich holds a permanent Investigator position from CONICET, currently at the International French-Argentine Center for Information and Systems Science (CIFASIS). Since 2008, Dr. Haimovich also holds an Adjunct Professor position at the School of Electronics Engineering, UNR. His research interests include nonlinear, switched and networked control systems.
\end{biography}

\end{document}